\documentclass[a4paper,11pt]{article}
\pdfoutput=1 
\usepackage{jheppub} 
\usepackage[T1]{fontenc} 
\usepackage{dcolumn}
\usepackage{bm}
\usepackage{mathrsfs}
\usepackage{tabularx}
\usepackage{amsmath}
\usepackage{slashed}
\usepackage{footmisc}
\newcommand{\mathsym}[1]{{}}

\newcommand{\baz}{\begin{array}{cc}}
\newcommand{\bad}{\begin{array}{ccc}}
\newcommand{\ba}{\begin{array}{c}}
\newcommand{\ea}{\end{array}}
\newcommand{\be}{\begin{equation}}
\newcommand{\ee}{\end{equation}}
\newcommand{\bea}{\begin{eqnarray}}
\newcommand{\eea}{\end{eqnarray}}

\newcommand{\bi}{\begin{itemize}}
\newcommand{\ei}{\end{itemize}}
\newcommand{\bmt}{\begin{pmatrix}}
\newcommand{\emt}{\end{pmatrix}}
\newcommand{\bt}{\begin{tabular}}
\newcommand{\et}{\end{tabular}}

\newcommand{\benu}{\begin{enumerate}}
\newcommand{\eenu}{\end{enumerate}}

\newcommand{\bav}{\begin{array}{cccc}}

\usepackage{amstext,amssymb}
\usepackage{amsmath}
\usepackage{graphicx}
\usepackage{xspace}
\usepackage{color}
\usepackage{slashed}
\usepackage{hyperref}
\global\long\def\d{\partial}
\title{\boldmath LHC signatures of sterile neutrinos in a Minimal Radiative Extended Seesaw framework }
\author[a]{Sudhanwa Patra}
\author[a]{\hspace*{-0.1cm}, Utkarsh Patel}
\author[a]{\hspace*{-0.1cm}, Purushottam Sahu}

\affiliation[a]{Department of Physics, Indian Institute of Technology Bhilai, Raipur 492015, India}

\emailAdd{sudhanwa@iitbhilai.ac.in}
\emailAdd{utkarshp@iitbhilai.ac.in}
\emailAdd{purushottams@iitbhilai.ac.in}

\abstract{The presence of small neutrino masses and flavour mixings can be accounted for naturally in various models about extensions of the standard model, particularly in the Seesaw Mechanism models. In this work, we present a minimally extended seesaw framework with two right-handed neutrinos, where the active neutrino masses are derived in the radiative regime. Using the framework it can be shown that within certain mass limits, the light neutrino mass term can approach a form that is similar to its form under type-I Seesaw Mechanism. Apart from this, we show that the decay width of right-handed neutrinos (produced through the decay of W boson in a particle collider) is short enough to cause a sufficiently long lifetime for the particles, thus ensuring an observable displacement in the LHC between the production and decay vertices. We comment on the fact that these displaced vertex signatures thus can serve as a means to verify the existence of these right-handed neutrinos in future experiments. Lastly, we line up the possibility of our future work where the vertex signatures of particles greater than the mass of W boson can be worked upon.    
}

\begin{document} 
\maketitle
\flushbottom

\section{Introduction}
The standard Model of Particle Physics is the best understanding of the universe that our scientists currently have. It successfully explains three of the four fundamental interactions happening in nature and its predictions have been tested to a very high degree of accuracy in various experiments. The work developed by Sheldon L. Glashow~\cite{Glashow:1961tr}, Steven Weinberg~\cite{Weinberg:1967tq}, and Abdus Salam~\cite{Salam:1968rm} in the 1960s and 1970s resulted in the so-called electroweak theory which together with Quantum chromodynamics(QCD), the theory of strong interactions, forms the Standard Model. The existence of all the SM fermions and gauge bosons were confirmed by 1995 with the discovery of the top quark at the CDF~\cite{CDF:1995wbb} experiment at the Tevatron at Fermilab. However, the particle responsible for the mechanism of mass for fundamental particles remained undetected for a long time. The long wait for the last missing piece of the Standard Model was finally over in 2012 with the discovery of Higgs Boson in the ATLAS~\cite{ATLAS:2012yve} and CMS~\cite{CMS:2012qbp} experiments(part of LHC). The second aim of LHC has been to discover signs of particles predicted by theories beyond the Standard Model. The motivation for these searches have come from numerous theoretical and experimental hints even if we ignore the fact that it doesn't unify gravity with other forces.\\

In the Standard Model, neutrinos are thought to be massless because the mechanism~\cite{Higgs:1964pj} responsible to produce fermion masses require fermions of both chiralities to be present in nature. But, right-handed chiral neutrinos were never found with direct evidence in any experiment. Hence, neutrinos were considered massless. But, oscillation experiments with atmospheric, solar, and reactor neutrinos have concretely depicted that neutrinos have non-zero masses. These conclusive discoveries made scientists put forward theories beyond the Standard Model. Another evidence for physics beyond the Standard Model comes from the mysterious form of matter known as Dark Matter. Firstly hinted by Fritz Zwicky in the 1930s~\cite{Zwicky:1933gu}, now evidently makes up about 26$\%$ of the energy density of the universe and there seems to be no viable Dark Matter candidate in the Standard Model particle sector. Thus, the phenomena of neutrino masses and dark matter are experimental hints for the necessity of an extension of the Standard Model.

The discovery that neutrinos have mass and they mix with each other has put before us another vital question to speculate over; whether they are Dirac or Majorana~\cite{Majorana:1937vz} particles. Even more intriguing is the theoretical origin of such a tiny mass and the mass hierarchy among them. An attractive mechanism for explaining small neutrino masses is the so-called see-saw mechanism proposed in 1979 by Murray Gell-Mann, Pierre Ramond and Richard Slansky~\cite{GellMann:1980vs} working in the U.S and independently by Tsutomu Yanagida~\cite{Yanagida:1980xy} of Tokyo University. The idea is to extend the Standard Model by adding right-handed neutrinos which are Majorana type particles with very heavy masses, possibly associated with large mass scale$(10^{14}\text{GeV})$ at which the three forces of the standard model unify~\cite{Mohapatra:1979ia,Schechter:1980gr}. The name see-saw for the mechanism comes from the fact that the lightness of left-handed neutrino mass is associated with the heaviness of the right-handed neutrino, rather like a flea and an elephant perched on either end of a see-saw. This relationship between the masses is incorporated in the mass matrix term of the neutrino lagrangian where the physical masses of the two different handed neutrinos are made to depend inversely on each other, so a heavy right-handed neutrino naturally explains the smallness of left-handed neutrino mass. As discussed above, the different seesaw mechanisms like type-I~\cite{Minkowski:1977sc,Yanagida:1979as,GellMann:1980vs,Mohapatra:1979ia}, type-II~\cite{Cheng:1980qt, Lazarides:1980nt,Magg:1980ut,Schechter:1980gr,Wetterich:1981bx} and others that appropriately explain this tiny mass, require them to be Majorana particles. 

The canonical seesaw mechanisms with the natural value of Dirac neutrino masses or the Right-handed neutrino masses constraints the seesaw scale around $10^{14}$GeV, which makes it impossible to probe into, because of the current collider limits or by any near-future planner experiments. Also, there are interesting indications for the possible existence of sterile neutrinos in the eV mass scale using the results of the LSND experiment ~\cite{LSND:1996ubh,LSND:2001aii} at LAMPF which are in favour of short-baseline ${\overline{\nu}}_{\mu} \rightarrow {\overline{\nu}}_{e}$ oscillations that require the existence of sterile neutrinos at eV mass scale. These motivations compel the researchers to think about alternative mechanisms for generating light neutrino masses. Some of the well known low-scale seesaw mechanisms available in the literature are radiative seesaw mechanism~\cite{Ma:2006km,Bajc:2005aq,Arbelaez:2016mhg}, extended see-saw mechanism~\cite{Krishnan:2020xeq}. The main idea of these alternatives is to bring down the seesaw scale to an experimentally verifiable energy range. Few recent attempts in the context of the low-scale extended seesaw are inverse seesaw ~\cite{Nomura:2019xsb,Pongkitivanichkul:2019cvm}, linear seesaw ~\cite{Nomura:2020opk,Borah:2018nvu}. Here, the Standard Model is minimally extended with the inclusion of two types of neutrinos.

In the work~\cite{Chianese:2018agp}, the authors have explored the Type-I seesaw with an addition of 2 Heavy Neutral Leptons(HNLs) and explored the constrains on allowed active-sterile mixing in experiments like SPS and SHiP~\cite{Bonivento:2013jag,SHiP:2015vad}. In the work~\cite{BhupalDev:2012jvh}, the authors have shown that light and super-light neutrino masses can be accommodated in a Minimal Radiative Inverse Seesaw(MRIS) scheme. Based on a similar line of work, we in our work derive type-I seesaw dominance starting with a Minimally extended radiative seesaw approach. In our scheme, we consider minimal extension of the Standard Model with right-handed neutrinos$(\nu_{R})$ and left-handed sterile neutrinos$(\nu_{sL})$ (commonly referred as $(\nu_{s})$). The Dirac mass term between left-handed SM neutrinos$(\nu_L)$ and right-handed neutrinos$(\nu_{R})$, mixing term between $\nu_R$ and $\nu_s$ $\&$ large lepton number violating Majorana mass term for $\nu_R$ are allowed while all other mass terms connecting $\nu_L-\nu_L$, $\nu_L-\nu_s$ and $\nu_s-\nu_s$ are either vanishing or very much suppressed. These vanishing or suppressed mass term values can be explained by imposing extra U(1)~\cite{Nomura:2020azp} or flavour symmetry~\cite{Rodejohann:2015hka}. We call our scheme as Minimally Extended Radiative Seesaw(MERS), as the scheme requires a minimal extension of the Standard Model with $\nu_R$ and $\nu_s$, while the light neutrino masses are explained via a radiative mechanism. One of the features of MERS is that the light neutrino masses cancel out to zero in the non-radiative regime and are thus generated in the radiative regime. An equally interesting feature that we analytically show in our work is the dominance of Type-I seesaw after the inclusion of radiative correction term (arising from the quantum loop effects) in the expression to calculate light neutrino masses. Towards the end, we comment on the low-energy constraints on MERS from particle physics, astrophysics and cosmological data. We also point out the clear distinction of MERS from inverse seesaw by explicitly showing that MERS allows large lepton number violations while the latter predicts the pseudo-Dirac nature of the sterile neutrinos suppressing the lepton number violating signals.\\
      
\section{Type-I seesaw and motivation for Extended Seesaw}
\label{sec:rsm}
%
%
\subsection{Type-I seesaw}
The type-I seesaw mechanism for light neutrino masses, popularly known as canonical seesaw was originally proposed by \cite{Minkowski:1977sc, Mohapatra:1979ia, Yanagida:1979as, Schechter:1980gr}. It is the simplest mechanism that explains the lightness of SM neutrino masses with the introduction of heavy right-handed neutrinos. The right-handed neutrinos being singlet under SM, are used to write down a Dirac mass term between SM neutrinos $\nu_L$ and right-handed neutrinos $\nu_R$ along with a bare Majorana mass term for right-handed neutrinos $\nu_R$.\\
The mass matrix for type-I seesaw in the $(\nu_L, \nu^c_R)$ basis is given by
\begin{align}
\mathcal{M}_\nu \ = \ 
\begin{pmatrix}
0   & M_D \\
M_D & M_R    
\end{pmatrix},
\end{align}
With the type-I seesaw approximated mass hierarchy $M_D \ll M_R$, the contribution to the light neutrino masses is given by
\begin{align}
\label{eq:type-I}
m_\nu \ \approx \ -\frac{M_D^2}{M_R} \, .
\end{align}
Here, the expression for heavy right-handed neutrino mass is $m_N = M_R$. The $\mathcal{O}(0.1)$eV scale light neutrinos requires $M_R$ to be of the order of $10^{14}$~GeV with natural choice of $M_D$ around $100$~GeV. The other outcome of the type-I seesaw is the mixing between left-handed and a right-handed neutrino as $V_{\ell N} \propto \sqrt{m_\nu/m_N}$. To detect right-handed neutrinos, we need their masses in the experimentally accessible range and also need their mixing to be large with active neutrinos. Considering $100$~GeV right-handed neutrinos, the left-right neutrino mixing is constrained from oscillation data as, $V_{\ell N} \leq 10^{-6} \left(m_\nu/\mbox{0.1\,eV} \right) \left(\mbox{100\,GeV}/m_N\right)$ which makes the type-I seesaw framework experimentally inaccessible in the current detector limit. This motivates researchers to think of alternative seesaw mechanisms where firstly the seesaw scale could be brought down to the experimentally verifiable range and also the mixing between left- and right-handed neutrinos could be made large. 
Thus, in the next subsection, we introduce the model framework by adding two types of neutrinos along with the known SM particles which allow us to overcome the above-discussed constraints.
\subsection{Extended seesaw structure}
The idea of extended seesaw is to allow the seesaw scale operative at experimentally accessible range and provide large light-heavy neutrino mixing. This is implemented with the introduction of two types of sterile neutrinos (denoted by $\nu_R$ and $\nu_S$). The extensive work on extended seesaw structure can be found in the literature~\cite{Ma:1989tz, Johnson:1986gt, Zhang:2011vh}. One of the highly promising framework among them is the minimal inverse seesaw~\cite{mohapatra:1986aw, Nandi:1985uh, mohapatra:1986bd}that comes with a mass matrix structure given as,
\begin{align}
\mathcal{M}_\nu \ = \ 
\begin{pmatrix}
0 & M_D & 0     \\
M^T_D & 0   & M_{RS}   \\
0 & M^T_{RS} & \mu_S 
\end{pmatrix},
\end{align}
With the mass hierarchy $\mu_S$, $M_D \ll M_{RS}$, the light neutrino mass expression in minimal inverse seesaw is given as
\begin{align}
\label{eq: mnu_minimalinverse}
m_\nu \ \approx \ -\mu_S \frac{M_D^2}{M_{RS}^2} \, .
\end{align}
and the heavy neutrino forms a pseudo-Dirac pair with the mass of the order of $M_{RS}$. The expression for light neutrino masses in inverse seesaw has a direct proportionality to the lepton number violating scale $\mu_S$ leading to suppressed LNV signals. To have large LNV signals, one can add a large Majorana mass term to this seesaw structure without affecting the light neutrino masses at tree level. However, one has to be careful with radiative correction terms for light neutrino masses mediated with the exchange of heavy neutrinos at loop as discussed in ref~\cite{Pilaftsis:1991ug, Dev:2012sg}.

Similarly, minimal linear seesaw has been explored within extended seesaw scheme with the following structure,
\begin{align}
\label{eq:linear}
\mathcal{M}_\nu = 
\begin{pmatrix}
0     & M_D & \mu_L \\
M^T_D   & 0   & M_{RS}   \\
\mu^T_L & M^T_{RS} & 0 
\end{pmatrix},
\end{align}
with $\mu_L$, $M_D \ll M_{RS}$. 

The light neutrino masses in this case is given by 
\begin{align}
\label{eq: mnu_minimallinear}
m_\nu \ \approx \ -\mu_L \frac{M_D^2}{M_{RS}^2}\,.
\end{align}
In the present work, we wish to discuss the extended seesaw model while emphasizing the radiative contribution to light neutrino masses only and its implications in the following section. Our work is different from the inverse and linear see-saw models because the mass matrix in our framework doesn't allow a Majorana mass term for sterile neutrino $\nu_s$ like in inverse see-saw or a mixing term between the SM left-handed neutrino and the sterile neutrino $\nu_s$ like in linear see-saw. The allowed mass term is a bare Majorana mass term for right-handed neutrino $\nu_R$ and a mixing term between $\nu_s$ and $\nu_R$. So , although the model is based on the extended scheme but is different in matrix structure when compared with linear and inverse models. A recent work on the phenomenology of KeV-scale sterile neutrino in a Minimally Extended see-saw(MES) framework seems interesting~\cite{Das:2019kmn}. MERS model in our work has a scope to allow for a eV-scale sterile neutrino, which has its effects in $\beta\text{-decay}$, neutrino-less double $\beta\text{-decay}$, and cosmology~\cite{Giunti:2019aiy}.
\section{Framework of Minimal Extended Radiative Seesaw (MERS)}
In our work, we have considered a minimal extension of SM with two types of neutrino fermions $\nu_R$ and $\nu_S$ for the implementation of the radiative seesaw mechanism. We denote generic right-handed neutrinos with $\nu_R$ while the left-handed sterile neutrinos as $\nu_S$. Since both the neutral fermions are completely singlets under the SM gauge group, one can write down all possible mass terms between these two states. However, we limit ourselves by writing only the mixing term between $\nu_R$ and $\nu_S$ along with the bare Majorana mass term for right-handed neutrinos. The Majorana mass term for $\nu_S$ is simply assumed to be zero or forbidden by invoking some symmetry. Also, we allow the generic Dirac mass term between SM neutrinos $\nu_L$ and $\nu_R$ while the interaction term between $\nu_L$ and $\nu_S$ is forbidden. 
\subsection{MERS Lagrangian}
The relevant leptonic Yukawa interaction terms for "{\em Minimal Extended Radiative Seesaw}" is given by
\begin{eqnarray}
\label{eq:TheModel}
	\mathcal{L}_\text{MERS} 
	&=&  \mathcal{L}_\text{SM} + i \, \overline{\nu}_{jR} \slashed{\d} \,\nu_{jR} 
	                           + i \, \overline{\nu}_{kS} \slashed{\d} \,\nu_{kS} \nonumber \\
	      &-& {[Y_D]}_{jk} \overline{\ell}_{jL} \widetilde{H} \nu_{kR}
	      - \frac{1}{2} {[M_R]}_{jk} \overline{\nu^c_{jR}} \nu_{kR} 
	      - {[M_{RS}]}_{jk} \overline{\nu_{jR}} \nu_{kS} 
	      +h.c.
	  \,
\end{eqnarray}
Here $\mathcal{L}_\text{SM}$ is the well known SM lagrangian, next two terms correspond to the kinetic energy terms for the right-handed and sterile neutrinos respectively. The fourth term represents generic Dirac Yukawa interaction between left-handed lepton doublet with right-handed neutrinos while the fifth term is for bare Majorana mass for $\nu_R$ and finally, the last term corresponds to mixing between right-handed neutrinos and sterile neutrinos. 
After electroweak symmetry breaking, the resulting neutral fermion mass matrix in the basis the basis $\left(\nu_L, \nu^c_R, \nu_S \right)$ is given by
\begin{equation}
\mathcal{M}_\nu= \begin{pmatrix}
              0                   & M_D   & 0  \\
              M^T_D                 & M_R   & M_{RS} \\
              0          & M^T_{RS}     & 0
                \end{pmatrix}
\label{eqn:numatrix-mers}       
\end{equation}
with $M_D = Y_D v/\sqrt{2}$ is the Dirac neutrino mass in which $v\equiv v_{\rm EW} = 174~$GeV is the VEV of 
electroweak symmetry breaking scale.

\begin{table}[h]
\begin{center}
\begin{tabular}{|c|c|c|}
	\hline
			& Field	& $ SU(2)_L\times U(1)_Y$		\\
	\hline
	\hline
	Fermions	& $Q_L \equiv(u_L, d_L)^T$			& $(\textbf{2},~ 1/6)$		\\
			& $u_R$							& $(\textbf{1},~ 2/3)$		\\
			& $d_R$							& $(\textbf{1},~-1/3)$		\\
			& $\ell_L \equiv(\nu_L,~e_L)^T$	& $(\textbf{2},~  -1)$		\\
			& $e_R$							& $(\textbf{1},~  -2)$		\\
			& $\nu_R$						& $(\textbf{1},~   0)$		\\
			& $\nu_S$						& $(\textbf{1},~   0)$		\\
	\hline
	Scalars	& $H$							& $(\textbf{2},~ 1/2)$	\\
	\hline
	\hline
\end{tabular}
\caption{The particle content of the Minimal Extended Radiative Seesaw (MRES) with large lepton number violation where SM is extended minimally with two types of neutral fermions $\nu_R$ and $\nu_S$. The third column displays the representation of the fields under the Electro-Weak gauge group.\label{tab:MRES}}
\end{center}
\end{table}

It has been shown in the appendix \ref{section:vanishing} that the light neutrinos are massless at tree level while both right-handed and sterile neutrinos have Majorana masses within the "{\em Minimal Extended Radiative Seesaw}" with mass hierarchy $M_R \gg M_{RS} > M_D$. After integrating out the heaviest right-handed neutrinos, the effective tree-level contribution to light active and sterile neutrinos ($\mathcal{M}^\prime_{\nu_L \nu_S}\equiv {\cal M}_{\rm eff}^{\rm tree}$) can be written as,
\begin{eqnarray}
V^\dagger {\cal M}_{\rm eff}^{\rm tree} V^* = \left(\begin{array}{cc}
{\bf 0}_3 & {\bf 0}_3 \\
{\bf 0}_3 & -M^T_{RS} M^{-1}_R M_{RS} 
\end{array}\right),
\label{eq:block}
\end{eqnarray}
where ${\bf 0}_3$ is a null matrix of order 3 and the $6\times 6$ unitary matrix  $V$ diagonalizes the light active and sterile neutrinos after right-handed neutrinos are already integrated out and is expressed in terms of a approximated mixing $\zeta$,~\cite{Grimus:2000vj}
(upto the order ot $\mathcal{O}(1/M_{RS})$) as,
\begin{eqnarray}
V = \begin{pmatrix}
         {\bf 1}_3- \frac{1}{2} \zeta \zeta^\dagger & \zeta \\
        -\zeta^\dagger & {\bf 1}_3- \frac{1}{2} \zeta^\dagger \zeta
         \end{pmatrix}
\label{eq:V}
\end{eqnarray} 
From~(\ref{eq:block}), it is straightforward to obtain 
\begin{eqnarray}
\zeta = M_D M^{-1}_{RS}\,.
\end{eqnarray} 

\begin{figure*}
 \begin{center}
 \includegraphics[width=0.45\textwidth]{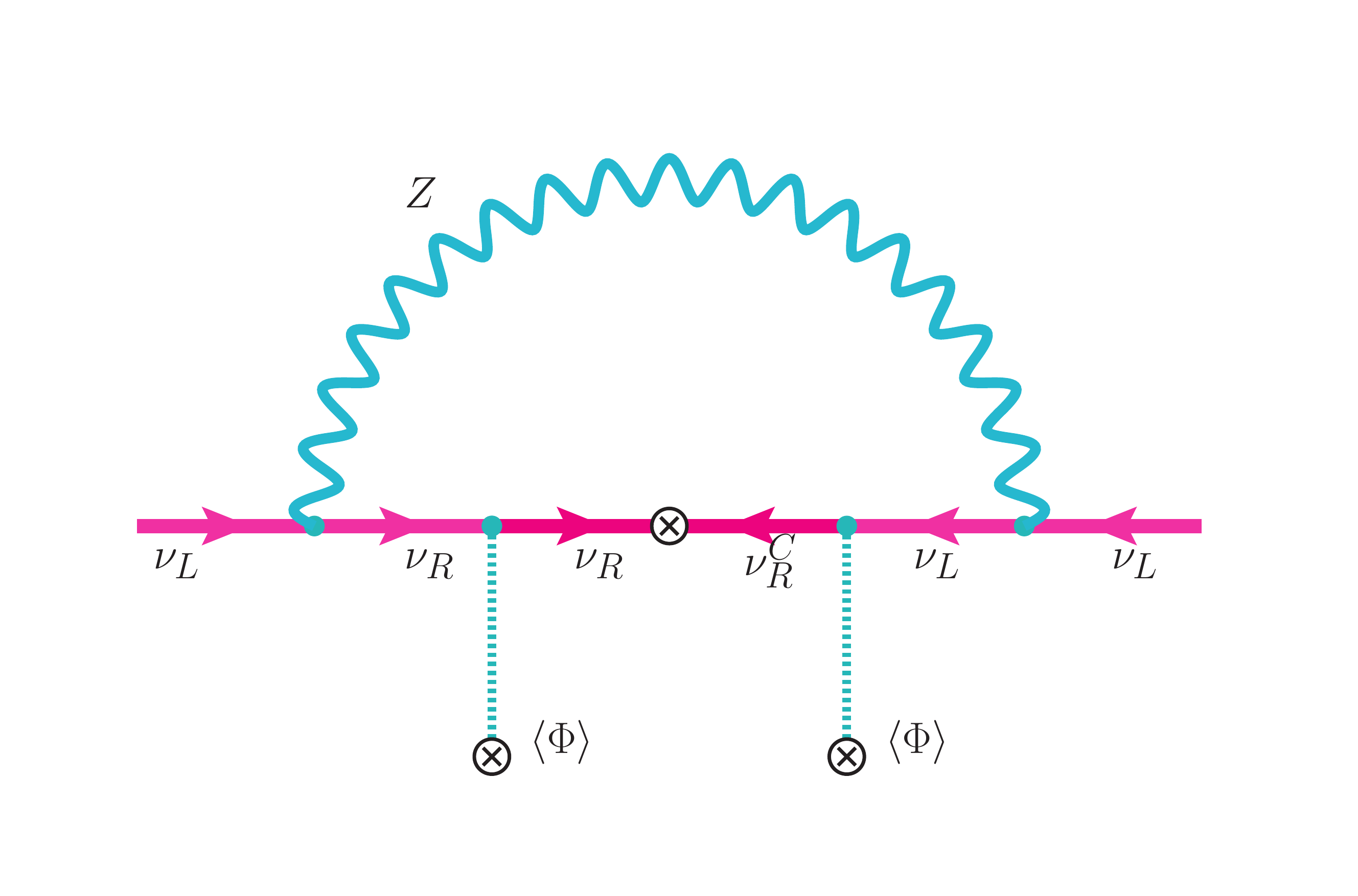}
 \includegraphics[width=0.45\textwidth]{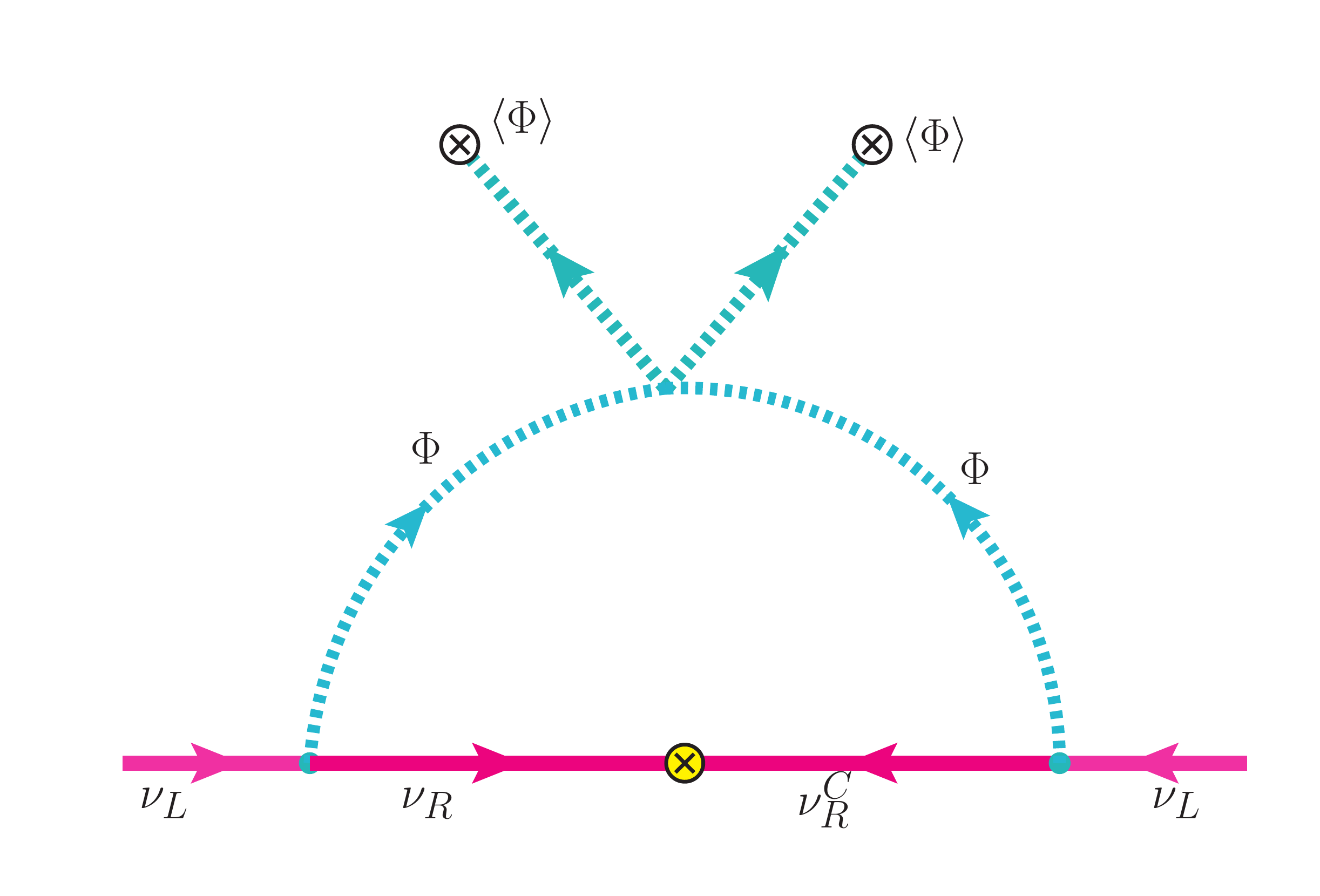}
 \caption{One loop contributions to light neutrino masses mediated by Z boson in the left panel and higgs boson in the right panel. These Feynman diagrams were created using the software \textit{JaxoDraw}, with the help of User Guide~\cite{Binosi:2008ig}.}
 \label{fig:figure1}
\end{center}
\end{figure*}
\subsection{Light neutrino masses via radiative mechanism}
It has been shown in the appendix~\ref{section:vanishing} that there are no tree level contributions to light neutrino masses in MERS scheme and type-I seesaw contribution is exactly cancelled out at leading order. The interesting feature of the model is that the heaviest right-handed neutrinos generate light active neutrino masses at one loop level via exchange of SM $W$ and $Z$ gauge bosons as shown in the Feynman diagram figure~\ref{fig:figure1}. The one loop contribution to light neutrino masses in the limit $\|M_R\| \gg \|M_D\|,\|M_{RS}\|$ has been shown in the Appendix section~\ref{section:Radiative} and is given by
\begin{eqnarray} \label{eq:lightest}
m^{\rm rad.}_\nu &\equiv & \Delta M \nonumber \\
 &&\hspace*{-0.8cm}= M^T_D \frac{\alpha_W}{16\pi
  m_W^2} M_R\left[\frac{m_H^2}{M_R^2-m_H^2{\bf 1}_3}\ln
  \left(\frac{M_R^2}{m_H^2}\right) +   
	\frac{3m_Z^2}{ M_R^2-m_Z^2{\bf 1}_3}\ln\left(\frac{
          M_R^2}{m_Z^2}\right)\right] M_D
\end{eqnarray}
After simplication by using the same seesaw approximations, it can be shown that the light active neutrino masses are governed by radiative seesaw mechanism and the same unitary mixing matrix diagonalizes it as,
\begin{eqnarray}
 &&m_\nu = \left({\bf 1}_3+\zeta\zeta^\dag\right)^{-1/2}\Delta M\left({\bf
  1}+\zeta^*\zeta^{\sf T}\right)^{-1/2} \nonumber \\
 && m_S \simeq  M_{RS} M_R^{-1} M^{T}_{RS} \label{eq:lighter}\\
 && \zeta = M_D M^{-1}_{RS} \nonumber \\
 && V_{6\times 6} = \begin{pmatrix}
                     \left({\bf 1}_3+\zeta\zeta^\dag\right)^{-1/2} & \zeta \\
                    -\zeta^\dagger   & \left({\bf 1}_3+\zeta^\dag \zeta\right)^{-1/2}
                    \end{pmatrix}
\end{eqnarray}

\subsection{Analysis of masses and mixing with 1-flavor case}
For illustrating that how one can get eV scale sterile neutrinos and large mixing with sub-eV scale active neutrinos, let us consider one flavor case where these matrices reduce to complex numbers and thus a subscript of "1" is used in the notation for neutrino masses to denote the same. Using relations given in~(\ref{eq:mformula})
and~(\ref{eq:m2}), the eigenvalues for one mass eigenstates of light active and sterile neutrino are given as 
\begin{eqnarray}
m_{{\nu}_{1}} &=& x_Rf(x_R) \frac{M^2_D}{M_R} \bigg(\frac{1}{1+|\zeta|^2} \bigg) \nonumber \\
&=& x_Rf(x_R) \frac{1}{M_R} \cdot M^2_{RS} \cdot \frac{M^2_D}{M^2_{RS}}  \bigg(\frac{1}{1+|\zeta|^2} \bigg) \nonumber 
\\
&=&\frac{x_Rf(x_R) M^2_{RS}}{M^2_{R}}\left(\frac{\zeta^2}{1+|\zeta|^2}\right)\;, \nonumber \\
m_{{s}_{1}} &=& \left( 1+|\zeta|^2 \right)^{1/2} M_{RS} M_R^{-1}M^{ T}_{RS} \left( 1+|\zeta|^2 \right)^{1/2} \nonumber \\
&=& 
\frac{M^2_{RS}}{M_R}(1+|\zeta|^2)\; ,
\end{eqnarray} 
and the mass eigenvectors are given by 
\begin{eqnarray}
  \label{eq:eigvec}
\left(\begin{array}{c}
n_1 \\ {n}_2 \end{array}\right) = {\cal V}^\dag\left(\begin{array}{c}
{\nu}_L \\ {\nu}_S \end{array}\right) 
= ({1}+|\zeta|^2)^{-1/2} \left(\begin{array}{c}
{\nu}_L-\zeta^* \cdot \nu_S \\
\zeta \cdot {\nu}_L + \nu_S
\end{array}\right)\; . 
\end{eqnarray}
\begin{itemize}
 \item $|M_{RS}|/M_R=10^{-4}$, one can have a sterile neutrinos of keV mass which can be dark matter candidate.
 \item $|M_{RS}|/M_R=10^{-5}$, we get eV scale sterile neutrino explanining LSND anomalies or have large impact on neutrino oscillation probabilities. 
\end{itemize}
A Recent Review on the sterile neutrino search experiments can be found in the reference~\cite{Seo:2020ehv}. 

\begin{table}
\begin{center}
\begin{tabular} {|c|c|c|c|c|c|c|}\hline
$M_D$(GeV) & $M_{RS}$(GeV) & $M_R$ & $m_\nu$(eV) & $m_S$(eV) & $m_N$(GeV)  & $C_{\rm loop}$ \\ \hline 
$0.8\times10^{-3}$ & $8 \times 10^{-3}$ & $200$  & $0.022$ & $320$ & $200$ & 0.00705\\
 \hline
$0.8\times10^{-3}$ & $8 \times 10^{-3}$ & $200$  & $0.025$ & $320$ & $200$ & 0.00705 \\
  \hline
$0.2 \times 10^{-3} $&$1 \times 10^{-3}$  &$200$  & $0.0014$  & $5$ &$200$    &$0.00705$  \\
  \hline
$0.2 \times 10^{-3} $&$1 \times 10^{-3}$  &$300$  &$0.00126$  & $3.33$ &$300$  & 0.00945  \\
  \hline
$0.3\times 10^{-3}$ &$1\times 10^{-3}$  & $1000$ & $0.0016$  & $1$  & $1000$   &  0.01788  \\
  \hline
$0.3\times 10^{-3}$ & $2\times 10^{-3}$ & $1000$ & $0.0016$ & $4$ &$1000$  & 0.01788  \\
\hline
\end{tabular}
\end{center}
\caption{Field representations of our LRSM scenario.  }
\label{tab:1}
\end{table}

\begin{table}
\begin{center}
\begin{tabular} {|c|c|c|c|c|c|}\hline
  $M_D$(GeV) & $M_{RS}$(GeV) & $M_R$(GeV) & $\nu_L-\nu_S$ mixing & $\nu_L-\nu_R$ mixing & $\nu_S-\nu_R$ mixing \\ \hline
 $0.8\times10^{-3}$ & $8 \times 10^{-3}$ & $200$  & $0.1$ & $4 \times 10^{-6}$ & $4 \times 10^{-5}$  \\
 \hline
  $0.8\times10^{-3}$ & $8 \times 10^{-3}$ & $200$  &$0.2$  &$1\times 10^{-6}$  &$5\times 10^{-6}$  \\
  \hline
 $0.2 \times 10^{-3} $&$1 \times 10^{-3}$  &$200$  & $0.2$ & $6.6667 \times 10^{-7}$ &$3.33 \times 10^{-6}$      \\
  \hline
 $0.2 \times 10^{-3} $&$1 \times 10^{-3}$  &$300$ & $0.2$ & $6.6667 \times 10^{-7}$ &$3.33 \times 10^{-6}$  \\
  \hline
 $0.3\times 10^{-3}$ &$1\times 10^{-3}$  & $1000$ &$0.3$  &$3\times 10^{-7}$  &$1\times 10^{-6}$  \\
  \hline
 $0.3\times 10^{-3}$ & $2\times 10^{-3}$ & $1000$ &$0.15$  &$3\times 10^{-7}$  & $2\times 10^{-6}$   \\
\hline
\end{tabular}
\end{center}
\caption{Field representations of our LRSM scenario.  }
\label{tab:1}
\end{table}
%
\subsection{Physical masses and mixing in MERS}
The detailed discussion on block diagonalization proceedue is presented in appendix~\ref{section:Radiative} for all neutrinos and the results are presented below as, ~\ref{eq:nuR}  ~\ref{eq:lighter} ~\ref{eq:lightest}
\begin{eqnarray}
&&m_\nu \equiv m^{\rm rad.}_\nu= M^T_D \frac{\alpha_W}{16\pi
  m_W^2} M_R\left[\frac{m_H^2}{M_R^2-m_H^2{\bf 1}_3}\ln
  \left(\frac{M_R^2}{m_H^2}\right) +   
	\frac{3m_Z^2}{ M_R^2-m_Z^2{\bf 1}_3}\ln\left(\frac{
          M_R^2}{m_Z^2}\right)\right] M_D \nonumber
  \label{eq:mnuformula}
\end{eqnarray}
and for the sterile and right-handed neutrinos, respectively, as
\begin{eqnarray}
  \label{eq:mSformula}
m_{S} &\simeq & M_{RS} M_R^{-1}M^{ T}_{RS}\; \\
m_N &\simeq& M_R 
  \label{eq:mNformula}
\end{eqnarray}
The physical masses for these neutral leptons are obtained by respective unitary mixing matrices as follows,
\begin{eqnarray}
 &&U^\dagger_\nu m_\nu U^*_\nu = \hat{m}_\nu = \mbox{diag}\left(m_{\nu_1}, m_{\nu_2}, m_{\nu_3} \right) \nonumber \\
 &&U^\dagger_S m_S U^*_S = \hat{m}_S = \mbox{diag}\left(m_{s_1}, m_{s_2}, m_{s_3} \right) \nonumber \\
  &&U^\dagger_N m_N U^*_N = \hat{m}_N = \mbox{diag}\left(m_{N_1}, m_{N_2}, m_{N_3} \right)
\end{eqnarray}
The full $9\times9$ mixing matrix within the allowed mass hierarchy $M_R \gg M_{RS} > M_D$ following the standard block diagonalization procedures is given by
\begin{eqnarray}
\mathcal{V}
&=& \begin{pmatrix}
U_{\nu}\left(1-\frac{1}{2}X X^{\dagger} \right)  & -U_{S}X &  Z U_N\\
   -U_{\nu}X^{\dagger}  & U_S \left(1-\frac{1}{2}X^{\dagger}X \right) & U_{N} Y \\
   \left(Z^\dagger X^\dagger X\right)U_\nu & -U_S Y^{\dagger} & U_N\left(1-\frac{1}{2}Y^{\dagger}Y \right)
\end{pmatrix}
\end{eqnarray}
Here the parameters used for mixing between different neutral fermion states are, $ X = M_{RS} M^{-1}_{R}$, $Y = M_{RS} M^{-1}_{R}$ and $Z=M_D M^{-1}_R$. 

\section{Deducing type-I dominance for light neutrino masses within MERS}
\label{sec:mers-typeI}
The light neutrino mass formula by radiative mechanism (while neglecting the supressed $\zeta$ factor), can be read as 
\begin{eqnarray}
m_\nu &= & M^T_D \frac{\alpha_W}{16\pi
  m_W^2} M_R\left[\frac{m_H^2}{M_R^2-m_H^2{\bf 1}_3}\ln
  \left(\frac{M_R^2}{m_H^2}\right) +   
	\frac{3m_Z^2}{ M_R^2-m_Z^2{\bf 1}_3}\ln\left(\frac{
          M_R^2}{m_Z^2}\right)\right] M_D
\end{eqnarray}
The series expansion of this radiative contribution to light neutrino masses upto order of ${\cal O}\left(1/M_R \right)$ is given by
\begin{eqnarray}
\hspace*{-0.5cm}m_\nu &= & \bigg(\frac{\alpha_W}{16\pi
  m_W^2} \bigg) M^T_D \bigg[ \frac{m^2_H}{M_R} \ln
  \left(\frac{1}{m_H^2}\right) + 2 \left( m_H^2 + 3 m^2_Z \right) \frac{\ln
  \left({M_R}\right)}{M_R} + \frac{3 m^2_Z}{M_R} \ln
  \left(\frac{1}{m_Z^2}\right) \bigg] M_D \nonumber \\
\end{eqnarray}
For large value of $M_R$, the second term in the square bracket vanishes \footnote{For large value of x, the limiting value of $\frac{\ln[x]}{x}$ vanishes as,
\begin{eqnarray}
  \lim_{x\to\infty}  \frac{\ln[x]}{x} = 0 \nonumber  \,.
\end{eqnarray}
} and the resulting light neutrino mass formula looks similar to type-I seesaw formula given as: 
\begin{eqnarray}
m_\nu &= & M^T_D M^{-1}_R M_D \bigg(\frac{\alpha_W}{16\pi
  m_W^2} \bigg) \bigg[ m^2_H \ln
  \left(\frac{1}{m_H^2}\right) +3 m^2_Z \ln
  \left(\frac{1}{m_Z^2}\right) \bigg] \nonumber \\
  &=& C_I \cdot M^T_D M^{-1}_R M_D = (-0.037) M^T_D M^{-1}_R M_D
\end{eqnarray}
Representative set of parameters and their allowed range is presented below:
\begin{eqnarray}
&&M_D = \mbox{1\,MeV}, \quad M_{RS} = \mbox{10\,GeV}, \quad M_{R} = \mbox{100\,GeV}\, \nonumber \\
&&\mathcal{V}_{}^{\nu S} \simeq 10^{-4}\,, \mathcal{V}_{}^{\nu N} \simeq 10^{-5}\,, \quad 
\mathcal{V}_{}^{S N} \simeq 10^{-1}\,
\nonumber \\
&&m_\nu \simeq {\mathcal O(0.1)}\mbox{eV}\,, \quad m_S \simeq {\mathcal O(1)}\mbox{GeV}, \quad m_N \simeq {\mathcal O(100)}\mbox{GeV}\,,
\end{eqnarray}
while another interesting sets of paremeter scan can be done to obtain eV scale sterile neutrinos as,
\begin{eqnarray}
&&M_D = \mbox{0.3\,MeV}, \quad M_{RS} = \mbox{1\,MeV}, \quad M_{R} = \mbox{100\,GeV}\, \nonumber \\
&&\mathcal{V}_{}^{\nu S} \simeq 0.3\,, \mathcal{V}_{}^{\nu N} \simeq 0.3 \times 10^{-5}\,, \quad 
\mathcal{V}_{}^{S N} \simeq 10^{-5}\,
\nonumber \\
&&m_\nu \simeq {\mathcal O(0.01)}\mbox{eV}\,, \quad m_S \simeq {\mathcal O(1)}\mbox{eV}, \quad m_N \simeq {\mathcal O(100)}\mbox{GeV}\,,
\end{eqnarray}

\begin{figure*}
 \begin{center}
 \includegraphics[width=0.45\textwidth]{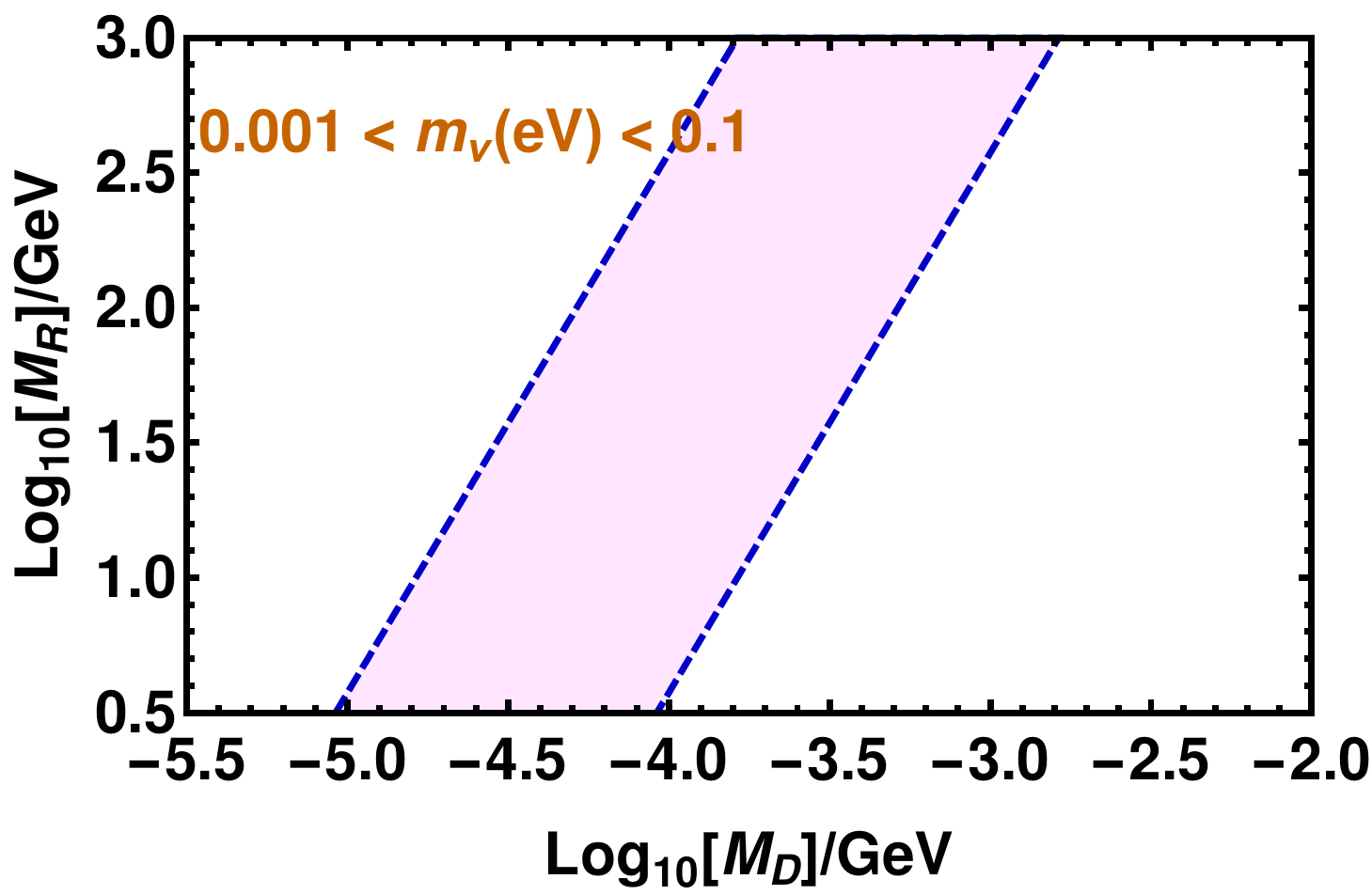}
  \includegraphics[width=0.45\textwidth]{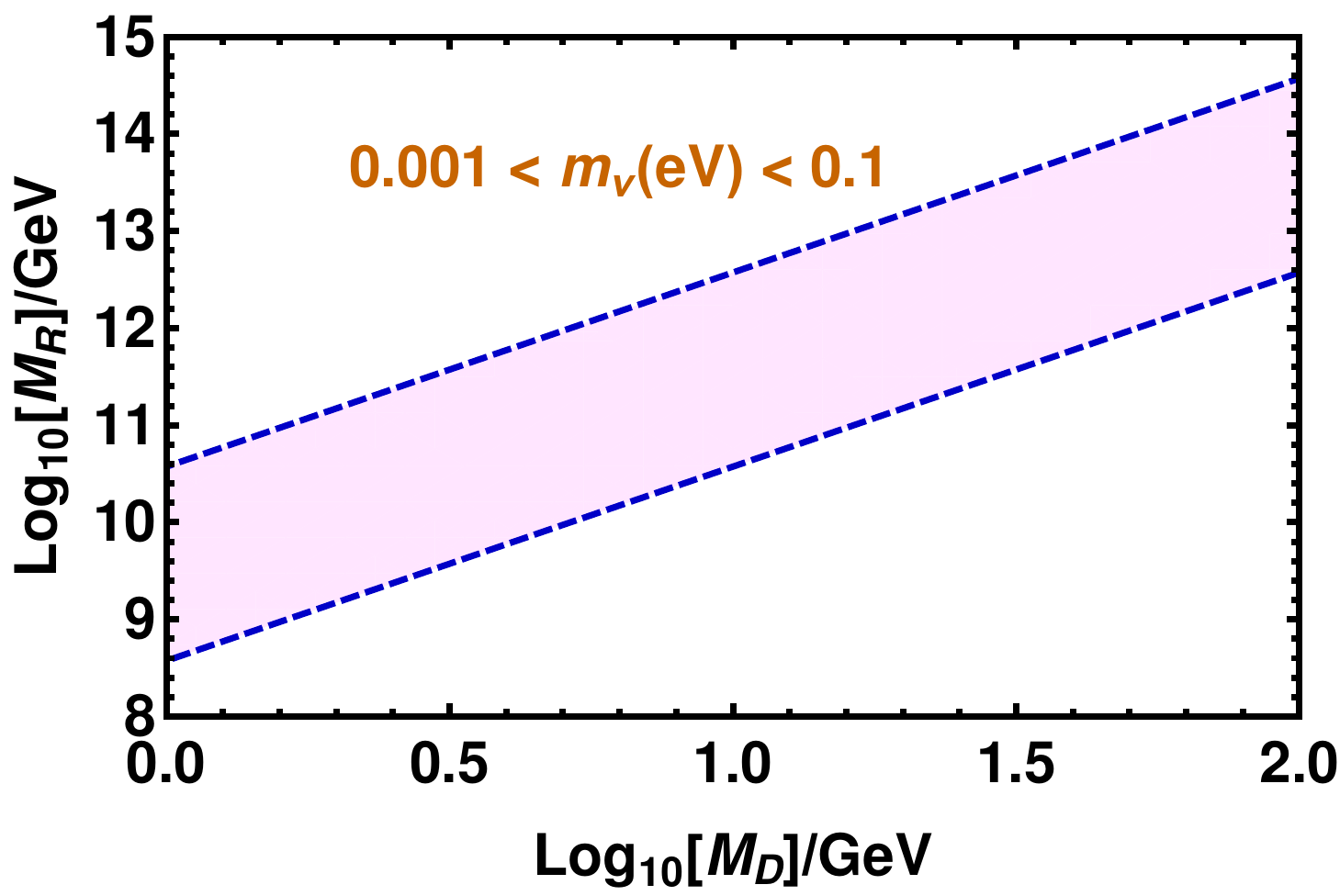}
 \caption{Allowed parameter space for Dirac neutrino mass $M_D$ and heavy right-handed neutrino Majorana mass $M_R$ demanding that the light neutrino mass will lie within the range, $0.001 < m_\nu\, \mbox{(eV)}< 0.1$ considering one generation scenario. The left-panel is for low mass range for $M_D$ and $M_R$ for achieving eV scale sterile neutrinos while the right-panel displays for large values of $M_D$ and $M_R$ suitable for thermal leptogenesis to account for matter-antimatter asymmetry.}
 \label{plot:mers-typeI}
\end{center}
\end{figure*}

The corelation between input parameters $M_D$ and $M_R$ for one generation of $\nu_L, \nu_R, \nu_S$ is displayed in Fig.\ref{plot:mers-typeI}.
\subsection{Relation between masses and mixings under type-I dominance} 
The mass matrices of the left-handed, right-handed and sterile neutrinos are proportional to each other,
\begin{eqnarray}
  &&m_\nu = -M^T_D M^{-1}_R M_D \, C_{\rm L} \propto  M^{-1}_R \,, \nonumber \\
  &&m_N = M_R \nonumber \\
  &&m_S = M^T M^{-1}_R M \propto  M^{-1}_R\, .
\end{eqnarray}

For three flavour scenario, one can consider Dirac neutrino mass matrix propertional to identity i.e, $M_D \propto k_d\, ({\bf 1}_{3 \times 3})$ (Here, $k_d$ is the proportionality constant with 'd' pointing to the Dirac nature of the term) and then the structure of $M_R$ will be decided by light neutrino masses and mixings. The complex symmetric right-handed neutrino mass matrix, $M_R$ is diagonalized as $\widehat{M_R} = U^\dagger_R M_R U^*_R$ and the light neutrino mass matrix is diagonalized as, 
$\widehat{m_\nu} = U^\dagger m_\nu U^*$. 
The interesting point here is that the unitary transformation matrices diagonalizing the active light left-handed and heavy right-handed neutrinos are themselves related in the following way
\begin{eqnarray}
 U_N = U^*_{\nu}=U^*_{PMNS}\,.
\end{eqnarray}
Similarly, ignoring the suppressed mixing factor $\zeta \propto M_D/M_{RS}$ and assuming $M_{RS} \propto k_{rs}\, ({\bf 1}_{3 \times 3})$(Here, $k_{rs}$ is the proportionality constant with 'rs' pointing to the mixing between heavy right-handed neutrino and sterile neutrino), one can verify that unitary matrices diagonalizing light sterile neutrinos $\nu_S$ and light active neutrinos are related as 
\begin{eqnarray}
 U_S = U_{\nu}=U_{PMNS}..
\end{eqnarray}
Thus, all the masses and mixing are fully deterministic for a fixed values of $k_d$ and $k_{rs}$.The mass eigenvalues for light as well as heavy neutrinos are related as
\begin{equation}
	M_i \propto 1/m_i \,,
\label{eq:typeI_dom_eigen_rel:a}
\end{equation}
where $M_i$ and $m_i$ are physical masses for light active and heavy Majorana neutrinos. 
Fixing $M_{\rm max}$( $M_3$ for NH and $M_2$ for IH), the mass relation can be expressed 
as

\begin{subequations}
\begin{eqnarray}
	M_{i} &= \frac{m_3}{m_{i}} M_{3},\text{ NH}, 
\label{eq:NH_massrel-typeII} \\
	M_{i} &= \frac{m_2}{m_{i}} M_{2},\text{ IH}.
\label{eq:IH_massrel-typeII} 
\end{eqnarray}
\end{subequations}

\newpage
\section{LHC signatures of sterile neutrinos}
Ever since the discovery of the Higgs boson in the LHC, no signs of new physics have been observed. This may hint towards the fact that current collider energy limits are not sufficient to probe new particles. But, an interesting scenario that can naturally explain these null LHC results can occur if new particles are SM singlet. Then, the particles cannot be produced or decay into known SM particles. Such particles may then be detected through new interactions. It may seem that a scenario like this one is even more challenging to test. But, if a new particle is stable for long enough, then it can leave a displaced vertex signature in the collider. Any electrically neutral particle which has a sufficiently long lifetime can display a signature of displaced vertex which is namely a vertex that is created by the decay of a particle that is located away from the point where the particle is produced. The displaced vertex signatures of known particles are well understood and thus an unknown vertex signature allow for a new particle's search with only a few events making this scenario a powerful probe to discover new particles. For the current status of displaced vertex searches at the LHC, one can refer to~\cite{ATLAS:2012av, CMS:2013czn, ATLAS:2012cdk}.\\
\noindent
Any signal leading to a lepton number violating interaction will serve as a hint for Majorana type right-handed neutral leptons. The collider signature accounting for this phenomenon would be an interaction producing same-sign dilepton plus two jets signal ($\ell^\pm \ell^\pm + 2j$), without missing energy. Several factors are affecting the final products of the lepton number violating interaction. The mixing between light-heavy neutrino and the mass of the sterile neutrino are the primary factors, but the type of production mechanism for these processes also contributes heavily. The decay half-life for heavy sterile neutrinos is dependent on their masses which in turn allow us to study and explore three distinct possibilities for the ($\ell^\pm \ell^\pm + 2j$) signal~\cite{Dib:2019ztn}.\\
In our framework, we have two sterile neutrinos namely $\nu_s$ and $\nu_R$. The mass hierarchy has been followed as: $\nu_s < \nu_R$. So, if the mass of any of these two sterile neutrinos is larger than the mass of W boson($M_W$), the neutrino decays immediately, and most probably into a charged lepton and two jets as shown in the figure~\ref{plot:mers-typeI}. But if the mass of any of the sterile neutrino is less than the mass of W boson that is in the regime of about $5~\text{GeV}$ up to $M_W$, then the neutrino will travel some distance before its decay which leads to a displaced vertex of leptons \cite{Helo:2013esa, Izaguirre:2015pga}. Thus, for neutrinos in this mass range, we expect the signal to be a promptly charged lepton and a displaced leptonic vertex. Based on the mass hierarchy in our model, we can have three distinct cases to study:
\begin{enumerate}
\label{enu:threecases}
 \item mass of $\nu_s$ is smaller than the mass of W boson i.e. $m_s < 80GeV$.
 \item mass of $\nu_R$ is smaller than the mass of W boson i.e. $m_R < 80GeV$.
 \item mass of $\nu_s$ is larger than the mass of W boson i.e. $m_s > 80GeV$.
\end{enumerate}

\noindent
The mass and the mixing formules for light active neutrinos, sterile neutrinos and heavy neutrinos in the MERS framework are given from the Appendix[\ref{eq:massformula}] and[\ref{eq:mixingformula}] as:
\begin{eqnarray}
m_{\nu} &=& \Delta M =  M_{D} M^{-1}_R x_R f\left(x_R \right) M^T_{D} \label{eq:mnu} \\
m_{S} &=& -M_{RS} M^{-1}_R M^T_{RS}  \label{eq:mS} \\
m_{N} &=& M_R \label{eq:mN}
\end{eqnarray}
and
\begin{eqnarray}
V_{\nu-S} &=&  M_{D} M^{-1}_{RS} \label{eq:Vmnu} \\
V_{S-N} &=& M_{RS} M^{-1}_R \label{eq:VmS}  \\
V_{\nu-N} &=& M_{D} M^{-1}_R \label{eq:VmN}
\end{eqnarray}

Table~\ref{tab:4} and~\ref{tab:5} shows the values of different parameters such as particle masses and mixings that come out naturally from the MERS framework and are within the ranges of above discussed three cases[\ref{enu:threecases}]. These table values have been obtained using the formulas[\ref{eq:mnu},\ref{eq:mS},\ref{eq:mN}] and [\ref{eq:Vmnu},\ref{eq:VmS},\ref{eq:VmN}] for one generation mass eigenstates case.  
\begin{table}
\begin{center}
\begin{tabular} {|c|c|c|c|c|c|c|c|}\hline
S. no. & $M_D$(GeV) & $M_{RS}$(GeV) & $M_R$(GeV) & $m_\nu$(eV) & $m_S$(GeV) & $m_N$(GeV)  & $C_{\rm loop}$ \\ \hline 
1 & $3\times10^{-2}$ & $3.1622 \times 10^{-2}$ & $10$  & $ \sim0.1$ & $1000 \times 10^{-9}$ & $\mathbf{10}$ & $10^{-4}$\\
 \hline
2 & $1\times10^{-2}$ & $1 \times 10^{2}$ & $1000$  & $\sim1$ & $\mathbf{10}$ & $1000$ & $10^{-2}$ \\
  \hline
3 & $1\times10^{-2}$ & $3 \times 10^{3}$ & $10000$  & $\sim0.1$ & $\mathbf{1000} $ & $\mathbf{10000}$ & $10^{-2}$\\
\hline
4 & $3\times10^{-3}$ & $3.1622 \times 10^{-3}$ & $1$  & $\sim0.03$ & $\mathbf{1000 \times 10^{-9}} $ & $\mathbf{1}$ & $10^{-6}$\\
 \hline
\end{tabular}
\end{center}
\caption{The table enlist 4 different sets of values of the various mass parameters of the MERS framework in GeV. In the last 3 columns, the output values of the three physical mass states has been written for the given mass parameter values. The output values have been calculated using the formulas given in equations \ref{eq:mnu},\ref{eq:mS} and \ref{eq:mN}.}
\label{tab:4}
\end{table}

\begin{table}
\begin{center}
\begin{tabular} {|c|c|c|c|c|c|}\hline
  $M_D$(GeV) & $M_{RS}$(GeV) & $M_R$(GeV) & $\nu_L-\nu_S$ mixing & $\nu_L-\nu_R$ mixing & $\nu_S-\nu_R$ mixing \\ \hline
 $3\times10^{-2}$ & $3.1622 \times 10^{-2}$ & $10$  & $0.948$ & $\mathbf{3 \times 10^{-3}}$ & $3.1622 \times 10^{-3}$  \\
 \hline
$1\times10^{-2}$ & $1 \times 10^{2}$ & $1000$  &$\mathbf{0.1 \times 10^{-3}}$  &$1\times 10^{-5}$  &$1\times 10^{-1}$  \\
  \hline
$1\times10^{-2}$ & $3 \times 10^{3}$ & $10000$  &$\mathbf{0.3 \times 10^{-5}}$  &$\mathbf{1\times 10^{-6}}$  &$3\times 10^{-1}$  \\
  \hline
 $3\times10^{-3}$ & $3.1622 \times 10^{-3}$ & $1$  & 0.948 &  $\mathbf{3.33 \times 10^{-3}}$ & $3.1622 \times 10^{-3}$  \\ 
  \hline
\end{tabular}
\end{center}
\caption{The table enlist 4 different sets of values of the various mass parameters of the MERS framework in GeV. In the last 3 columns, the output values of mixing between the three physical mass states has been written for the given mass parameter values. The output values have been calculated using the formulas given in equations \ref{eq:Vmnu},\ref{eq:VmS} and \ref{eq:VmN}. }
\label{tab:5}
\end{table}
 \subsection{Cases where mass of sterile neutrino(N) is less than the mass of W boson }

\begin{figure*}
 \begin{center}
 \includegraphics[width=0.60\textwidth]{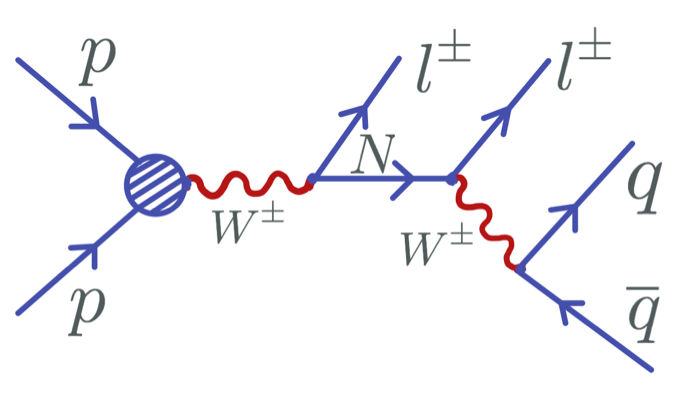}
 \caption{Feynamn Diagram depicting the delayed vertex signal corresponding to the case where mass of sterile neutrino( $\nu_s$) is less than the mass of W boson($W^+$). This Feynman Diagram along with the diagrams in figure \ref{plot:Fey2}, \ref{plot:Fey3} and \ref{fig:figure3} have been created using the software \textit{Notability} on \textit{MacOS}, with the help of its User Guide. The details of which have been added in the footnote:[\ref{fnlabel}].}
 \label{plot:Fey1}
\end{center}
\end{figure*}

\begin{figure*}
 \begin{center}
 \includegraphics[width=0.60\textwidth]{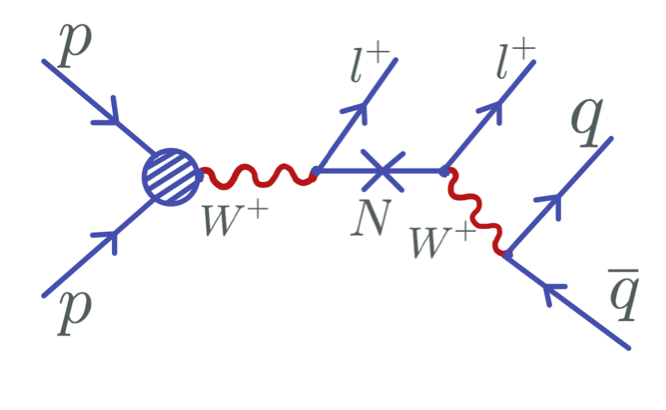}
 \caption{Feynamn Diagram depicting the delayed vertex signal corresponding to the case where W boson($W^+$) decays into a Majorana-type sterile neutrino( $\nu_s$) plus an antimuon($\mu^{+}$).}
 \label{plot:Fey2}
\end{center}
\end{figure*}

\begin{figure*}
 \begin{center}
 \includegraphics[width=0.60\textwidth]{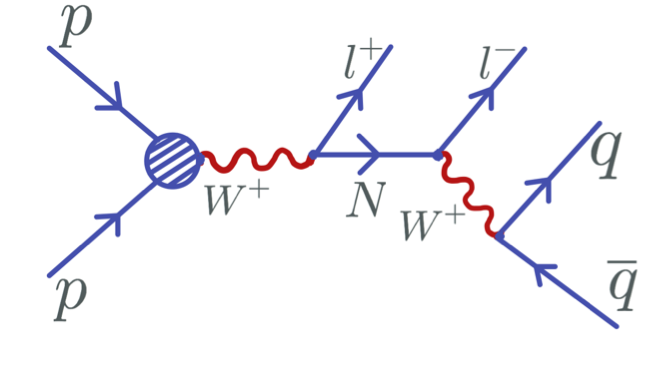}
 \caption{Feynamn Diagram depicting the delayed vertex signal corresponding to the case where W boson($W^+$) decays into a Dirac-type sterile neutrino( $\nu_s$) plus a muon($\mu^{-}$).}
 \label{plot:Fey3}
\end{center}
\end{figure*}

The MERS framework in its neutral lepton sector consists of the usual three generations of left-handed SM neutrinos along with the introduction of three families of two right-handed sterile neutrinos $\nu_s$ and $\nu_R$.
The added sterile leptons are called so,
because other than the mixing composition with each other and the usual left-handed neutrinos, these  
do not directly interact with any other SM particles.
At the LHC, sterile neutrinos with masses around $5\sim 20 GeV$ will be mainly produced from the decay of on-shell $W$ bosons. Thus, when the mass of the sterile neutrino is less than the mass of W bosons, then W boson decays into the sterile neutrino and a lepton, which thereafter decays into a lepton plus other particles like pions. This scenario can be referred from the figure~\ref{plot:Fey1}. In reference from~\cite{Dib:2019ztn}, it is inferred that most promising modes should be $W^+\rightarrow \mu^+N$ followed by a displaced decay either into $N \rightarrow \pi \mu^+$,$2\pi \mu^+$ or $3\pi \mu^+$ for a Majorana sterile neutrino as can be seen from the figure~\ref{plot:Fey2}, or $W^+\rightarrow \mu^+N$ followed by $N \rightarrow \pi \mu^-$,$2\pi \mu^-$ or $3\pi \mu^-$ for a Dirac neutrino as can be seen from the figure~\ref{plot:Fey3}.
 The decay rate $W^+\to \ell^+ N$ has been derived in the reference~\cite{Dib:2019ztn} by calculating the branching fraction of $W^+\to \ell^+ N$ concerning the total decay width of the W boson: 
\begin{align}
  {\cal B}(W^+ \to \ell^+ N) \equiv \frac{\Gamma(W^+\to \ell^+ N)}{\Gamma_W} =  \frac{G_F}{\sqrt{2}} \frac{M_W^3}{12\pi \Gamma_W} |V_{\ell N}|^2 \left( 2+\frac{m_N^2}{M_W^2}\right)\left( 1- \frac{m_N^2}{M_W^2}\right)^2,
\end{align}
Here, $\Gamma_W \simeq 2.085$ GeV is the total decay width of the $W$ boson~\cite{Patrignani:2016xqp}.

From here, the sterile neutrino $N$ can decay in several modes, depending on its mass. The case to look for is when the sterile neutrino decays into pions, namely $N \to \pi^\mp \ell^\pm$, $N \to \pi^0 \pi^\mp \ell^\pm$  and $N \to \pi^\mp \pi^\mp \pi^\pm \ell^\pm$. Both charged modes will occur for a Majorana $N$, while for a Dirac $N$,
only the mode where $N$ decays into a negatively charged lepton will be produced as shown in the table[\ref{tab:7}]

\begin{table}
\begin{center}
\begin{tabular}{c|c}
 
For a Majorana $N$: & For a Dirac $N$:\\ 
\hline
$\boldsymbol{N \to \pi^{\mp} \ell^{\pm}}$ & $\boldsymbol{N \to \pi^+ \ell^-}$ \\
$\boldsymbol{N \to \pi^0 \pi^{\mp} \ell^{\pm}}$ & $\boldsymbol{N \to \pi^0 \pi^+ \ell^-}$ \\
$\boldsymbol{N \to \pi^{\mp} \pi^{\mp} \pi^{\pm} \ell^{\pm}}$ & $\boldsymbol{N \to \pi^+ \pi^+ \pi^- \ell^-}$ 
\label{tab:7}

\end{tabular}
\end{center} 
\caption{Table depicting Decay Modes for neutrinos}
\end{table}

With the expressions described above we can estimate the exclusive semileptonic decay rates of $N$ into $\pi \ell$, $2\pi \ell$ and $3\pi\ell$, for a neutrino $N$ with mass in the range 5 to 20 GeV, produced at the LHC in the process $W \to \ell N$.

Reference~\cite{Dib:2019ztn} introduce a term called ``Canonical'' decay rates for $N \to \pi^- \ell^+$, $N \to \pi^0 \pi^- \ell^+$, $N \to \pi^- \pi^- \pi^+ \ell^+$ and ``canonical'' branching ratios for the full processes $W^+\to \ell^+\ell^+ n \pi$ which are basically the decay rates and branching ratios with all mixing factors $|V_{\ell N}|$ removed. To obtain the actual values, the ``canonical''  values must be multiplied by the factor
 $|V_{\ell N}|^2$, and $|V_{\ell N}|^4/\sum_{l_i} |V_{\ell_i N}|^2$, respectively.

The decay modes with final leptons as muons instead of electrons or tau particles are the most promising decay modes~\cite{Dib:2018iyr}.

The expected number of $W\to N \ell \to n\pi +\ell \ell $ events at the LHC,
%
or equivalently the minimal value of the lepton mixing element that would generate 5 events or more, for a benchmark value of
$m_N= 10$~GeV is given by~\cite{Dib:2019ztn}: 
\[
{\cal N}_W\times {\cal B}(W^+\to n \pi  +\mu^+ \mu^+)  \times \frac{|V_{\mu N}|^4}{\sum_{\ell} |V_{\ell N}|^2 }
\]

At this mass range, a canonical branching ratio is given as~\cite{Dib:2019ztn}:
\begin{align}
& {\cal B}(W^+\to \pi^- \mu^+ \mu^+, \pi^0\pi^- \mu^+ \mu^+, \pi^+\pi^-\pi^-\mu^+ \mu^+,\pi^0\pi^0\pi^-\mu^+ \mu^+)\nonumber\\
& \hspace{36pt} \equiv {\cal B}(W^+\to n\pi +\mu^+\mu^+)
 \approx 8\times 10^{-4}.
\end{align}

 \footnote[2]{Ginger Labs, Inc. (2021). \textit{Notability: User Guide}. Retrieved from \url{http://eisdedtechs.weebly.com/uploads/1/3/1/9/13198293/notability_user_guide_p1-8.pdf} \label{fnlabel}}
 
According to
Ref.~\cite{Aad:2016naf}, at the end of the LHC Run II one may expect a sample of ${\cal N}_W \sim 10^9$ $W$ decays.
Therefore, to obtain more than 5 events, we must have:
\[
{\cal N}_W\times {\cal B}(W^+\to n \pi  +\mu^+ \mu^+)  \times \frac{|V_{\mu N}|^4}{\sum_{\ell} |V_{\ell N}|^2 }  > 5 ,
\]
which implies $|V_{\mu N}|^2 \gtrsim 6.2\times 10^{-6}$, provided other mixing elements are smaller. If instead, all mixing elements are comparable, then this lower bound increases by a factor 3, i.e.  \hbox{$|V_{e N}|^2,
\, |V_{\mu N}|^2, \, |V_{\tau N}|^2 \gtrsim 1.9 \times 10^{-5}$}. Now, following the framework of MERS, the table ~\ref{tab:4} and~\ref{tab:5} shows the two sterile neutrino candidates $\nu_s$ and $\nu_R$ where the mass of $\nu_R$ and coupling between $\nu_l$ and $\nu_R$ in the first row and the mass of $\nu_s$ and coupling between $\nu_l$ and $\nu_s$ in the second row, are within the bounds required to have 5 or more significant events at the LHC, respectively. These results are in ideal conditions,  with no cuts or backgrounds.

One last important point in the observability of these processes is the long lifetime of $N$, which would cause an observable displacement in the detector between the production and decay vertices of $N$. This displacement will drastically help reduce the possible backgrounds. For a sterile neutrino $N$ with mass in the range 5 GeV to 20 GeV, the total width can be estimated as~\cite{Dib:2015oka, Drewes:2019fou}:
\begin{equation}
\Gamma_N \sim \frac{G_F^2 m_N^5}{12 \pi^3} \sum_\ell |V_{\ell N}|^2.
\label{nwidth}
\end{equation}
We see from the above-written expression that, lighter the mass of $N$ and smaller the mixing $|V_{\ell N}|^2$ are, the longer $N$ will live. Using the current upper bound $|V_{\ell N}|^2\sim 19 \times 10^{-5}$, the characteristic displacement $\tau c  \equiv \hbar c/\Gamma_N$  is in the range $\sim 20 \ \mu m$ to 20 $mm$ (for 20 GeV and 5 GeV respectively). For smaller $|V_{\ell N}|^2$, the displacements will be proportionally larger. Moreover, the
relativistic $\gamma$ factor will increase the displacement as well.

\subsection{Case where masses of sterile neutrinos(N) are greater than the mass of W boson }
The case in this sub-section is something that we in our present work are not interested in. However, LHC searches for sterile neutrino masses greater than 100GeV~\cite{ATLAS:2015gtp} are based on inclusive processes $pp\rightarrow W^*X$, $W^*\rightarrow l^{\pm}l^{\pm}jj$ as shown in the references~\cite{Keung:1983uu,delAguila:2007qnc}. Taking these processes as a possibility, we can hope to extend our model in future works to include displaced vertex signals from processes involving decays of particles into neutrinos having masses beyond 100GeV. Then, further studies and implications of those heavier neutrinos in the field of astrophysics, cosmology and collider physics can be subsequently explored. 
\section{Conclusion}
\label{sec:conclusion}

The name "{\em Minimal Extended Radiative Seesaw}" of our framework is justified as the SM is extended minimally with $\nu_R$ and $\nu_S$ to accommodate light neutrino masses via a radiative mechanism in a see-saw type structure while allowing for large lepton number violation. By using a minimally extended radiative see-saw framework, we in this work have shown that with an equal number of $\nu_L$, $\nu_R$ and $\nu_S$, the canonical type-I seesaw contribution $m^{I}_\nu = -M^T_D M^{-1}_R M_D$ is exactly cancelled out and the active neutrino masses are thus generated at 1-loop level via the mediation of SM gauge bosons ($W$ and $Z$). However, this result is well known previously. Using these results, we set the mass ranges and mixings of three neutrinos in a tabular structure~\ref{tab:1} to show that our model has the potential to explain eV scale sterile neutrinos leading to new physics contributions to neutrino oscillation experiments via non-standard neutrino interactions and neutrinoless double beta decay. A recent study on the role of eV-scale sterile neutrinos in particle physics and cosmology can be found in the references~\cite{Kang:2019xuq,Dasgupta:2021ies}.\\
The charge-current interactions in flavor basis and mass basis by expressing $\nu_{\alpha}$ as a linear combination of all massive states introduced in our work, leads to modified charge-current interactions as given in the equation~\ref{eq:ccLagrangian}. This can have interesting consequences to the contribution of g-2 anomaly for a muon, which can be explored well in future works.\\
In section 6, the possible production mechanism of right-handed neutrinos from the decay of W bosons have been discussed and it has been shown that the framework can be testified by tracing the signatures of lepton jets at the Large Hadron Collider (LHC) and at the future collider experiments produced due to the displaced vertices generated from the decay of right-handed neutrinos of masses less than 100GeV which themselves are produced from the decay of W bosons (at a location different from their decay point into those lepton pairs). The possibility of displaced vertex signals for the production of neutrinos heavier than 100GeV have been left out for future works.  


\begin{figure*}
 \begin{center}
 \includegraphics[width=1.0\textwidth]{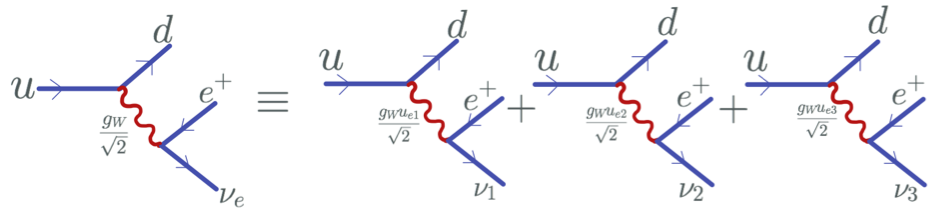}
 \caption{Illustration of charge-current interactions in flavor basis and mass basis for first generation of leptons. There is only one diagram for flavor basis while there are three diagrams in the mass basis arising from the neutrino oscilations where each neutrino flavor is expressed as a linear combination of three mass eigenstates $\nu_{\alpha}=\sum U_{\alpha i} \nu_{i}$}.
 \label{fig:figure3}
\end{center}
\end{figure*}

\newpage
\section{Appendix}
\appendix
\vspace*{-0cm}
\section{Derivation of neutrino masses and mixings in Minimal Extended Radiative Seesaw (MERS)}
\label{app:lrsm}
\subsection{Mass Matrix for MERS}

The relevant Yukawa terms leading to Minimal Radiative Extended Seesaw (MERS) structure is given by
\begin{eqnarray}
\mathcal{L}_{\tiny \rm MERS} &=& \mathcal{L}_{M_D} + \mathcal{L}_{M_{RS}} + \mathcal{L}_{M_R} \nonumber \\
\mathcal{L}_{M_D} &=& - \sum_{\alpha, \beta} \overline{\nu_{\alpha L}} [M_D]_{\alpha \beta} \nu_{\beta R} \mbox{+ h.c.}\nonumber \\
\mathcal{L}_{M_{RS}} &=& - \sum_{\alpha, \beta} \overline{\nu_{\alpha S}} [M_{RS}]_{\alpha \beta} \nu_{\beta R} \mbox{+ h.c.}
\nonumber \\
\mathcal{L}_{M_{R}} &=& - \frac{1}{2} \sum_{\alpha, \beta} \overline{\nu^c_{\alpha R}} [M_{R}]_{\alpha \beta} \nu_{\beta R} \mbox{+ h.c.}
\end{eqnarray}
The flavour states for light active neutrinos $\nu_{\alpha L}$, sterile neutrinos $\nu_{\beta S}$ and right-handed neutrinos $\nu_{\gamma R}$ are defined as follows
\begin{eqnarray}
 \nu_{\alpha L} = \begin{pmatrix}
                   \nu_{eL} \\ \nu_{\mu L} \\ \nu_{\tau L}
                  \end{pmatrix}\, , \quad 
 \nu_{\beta S} = \begin{pmatrix}
                   \nu_{S_{1 L}} \\ \nu_{S_{2 L}} \\ \nu_{S_{3 L}}
                  \end{pmatrix}\, , \quad 
  \nu^c_{\gamma R} = \begin{pmatrix}
                   \nu^c_{N_{1 R}} \\ \nu^c_{N_{2 R}} \\ \nu^c_{N_{3 R}}
                  \end{pmatrix}\, .
\end{eqnarray}
Similarly, the mass states for neutral fermions are given by
\begin{eqnarray}
 \nu_{i L} = \begin{pmatrix}
                   \nu_{1L} \\ \nu_{2 L} \\ \nu_{3L}
                  \end{pmatrix}\, , \quad 
 S_{j L} = \begin{pmatrix}
                   S_{1 L} \\ S_{2 L} \\ S_{3 L}
                  \end{pmatrix}\, , \quad 
  N^c_{k R} = \begin{pmatrix}
                   N^c_{1 R} \\ N^c_{2 R} \\ N^c_{3 R}
                  \end{pmatrix}\, .
\end{eqnarray}

The full $9\times 9$ neutral lepton mass matrix for MERS can be re-expressed in the the modified flavour basis $\left( \nu_L, \nu_S, \nu^c_R\right)$ as:
\begin{eqnarray}
\mathcal{M}_{\tiny \rm MERS}= \begin{pmatrix}
 0      & 0           & M_D  \\
 0      & 0           & M_{RS} \\
 M^T_D  & M^T_{RS}    & M_R
\end{pmatrix}
\label{eq:numatrix-complete-modified}
\end{eqnarray}
Here $M_D$, $M_{RS}$ and $M_R$ are Dirac mass term connecting $\nu_L-\nu_R$, mixing term between $\nu_R-\nu_S$ and Majorana mass term for $\nu_S$, respectively. All the entries are $3\times3$ order matrices and the mass hierarchy considered is given by
$$M_R \gg M_{RS} > M_D\, .$$
The diagonalization of MERS mass matrix is done by going from flavour basis of all neutral leptons to mass basis. The change of flavor basis to mass basis is achieved by an $9\times 9$ unitary mixing matrix as follows,
\begin{eqnarray}
 &&\mid \Psi \rangle_f = \mathcal{V}^*_{9 \times 9} \mid \Psi \rangle_m\\
 &\mbox{or,}& \begin{pmatrix}
               \nu_{\alpha L}\\ \nu_{\beta S}\\ \nu^c_{\gamma R}
              \end{pmatrix}
              = 
              \begin{pmatrix}
               \mathcal{V}_{\alpha i}^{\nu \nu} & \mathcal{V}_{\alpha j}^{\nu S} & \mathcal{V}_{\alpha k}^{\nu N}\\\mathcal{V}_{\beta i}^{ S \nu} & \mathcal{V}_{\beta j}^{S S} & \mathcal{V}_{\beta k}^{S N}\\\mathcal{V}_{\gamma i}^{N \nu} & \mathcal{V}_{\gamma j}^{N S} & \mathcal{V}_{\gamma k}^{N N}
               \end{pmatrix}               
               \begin{pmatrix}
               \nu_{i}\\ S_{j}\\ N^c_{k}
              \end{pmatrix} \, \\
      &&   \mathcal{V}^{\dagger} \mathcal{M_{\rm MERS}} \mathcal{V}^* 
      = \mathcal{\widehat{M}}_{\rm MERS} \nonumber \\
      &&\hspace*{2.2cm} = \mbox{diag} \left(m_{i},m_{S_j},m_{N_k} \right) \nonumber \\
      &&\hspace*{2.2cm} = \mbox{diag} \left(m_{1},m_{2},m_{3},m_{S_1},m_{S_2},m_{S_3},m_{N_1},m_{N_2},m_{N_3} \right)
\end{eqnarray}
We actually need to derive the form $\mathcal{V}_{9 \times 9}$ and $\mathcal{\widehat{M}}_{\rm \tiny MERS}$ 
using the allowed mass hierarchy. Using the seesaw approximations, we can write down the $\mathcal{V}_{9 \times 9}$ in terms of unitary mixing matrix $\mathcal{U}$ and approximate block diagonalized matrices $\mathcal{W}_1$ and $\mathcal{W}_2$ as,
\begin{eqnarray}
 \mathcal{V}_{9 \times 9} = \mathcal{W}_1 \cdot \mathcal{W}_2 \cdot \mathcal{U}_{9 \times 9}
\end{eqnarray}

\subsection{Exact cancelation of type-I seesaw and vanishing light neutrino masses-}
\label{section:vanishing}
We illustrate the proof of vanishing light neutrinos masses for SM neutrinos $\nu_L$ in the presence of two types of sterile neutrinos $\nu_R, \nu_{SL}$ with the assumption that the Majorana mass terms are of the form $\overline{\nu^C_L} \nu_L$, $\overline{\nu^C_S} \nu_{SL}$ and $\overline{\nu^C_L} \nu_{SL}$. 

\noindent \\
{\bf \underline{First Seesaw Approximation with rotation matrix $\mathcal{W}_1$:-}}\\
We can map the MERS mass matrix, $\mathcal{M}_{\tiny \rm MERS}$ given in eq.(\ref{eq:numatrix-complete-modified}) to generic type-I seesaw analysis~\cite{Mohapatra:1979ia}  as,
\begin{eqnarray}
\mathcal{M}^{\prime}_{\tiny \rm MERS} &=& \begin{pmatrix}
              {\bf 0}_{6 \times 6} & \mathcal{M}^T_{D}  \\
              \mathcal{M}_{D} & \mathcal{M}_{R}
             \end{pmatrix} \, , \nonumber \\
 \mbox{where,} && \mathcal{M}_{D}=  \begin{pmatrix}
                                  M^T_D & M^T_{RS}
                                 \end{pmatrix}\, , \quad  
                                 \mathcal{M}_{R} = M_R\,. 
\end{eqnarray}
As usual case of type-I seesaw case, we need to integrate out the heaviest state $\nu_R$ using the allowed mass hierarchy $| \mathcal{M}_{R} | \gg   |\mathcal{M}_{D}|$. Applying the first seesaw approximations $| \mathcal{M}_{R} | \gg   |\mathcal{M}_{D}|$, we get the resulting block diagonalized MERS mass matrix as,
\begin{eqnarray}
&&\mathcal{W}^T_1 \mathcal{M}^{}_{\tiny \rm MERS} \mathcal{W}_1=\mathcal{M}^{BD}_{\tiny \rm MERS} \nonumber \\
    && \mathcal{M}^{BD}_{\tiny \rm MERS} = \begin{pmatrix}
                                        M^{\rm light}_{MERS}        & {\bf 0}_{3 \times 6} \\
                                        {\bf 0}_{3 \times 6}  & M^{\rm heavy}_{MERS}
                                       \end{pmatrix} \nonumber \\
    &&\mathcal{W}_1=\begin{pmatrix}
             \sqrt{1-\mathcal{B}\mathcal{B}^{\dagger}} & \mathcal{B}\\
             -\mathcal{B}^{\dagger} & \sqrt{1-\mathcal{B}^{\dagger}\mathcal{B}}
            \end{pmatrix}
\end{eqnarray}
The block diagonalized mass matrix gives effective mass matrix in the basis $\left(\nu^{bd}_L, \nu^{bd}_{S} \right)$, which can be shown as: 
\begin{eqnarray}
M^{\rm light}_{MERS} &\equiv& M^{\rm Eff}_{MERS} =  - \mathcal{M}^T_{D} \mathcal{M}^{-1}_{R} \mathcal{M}_{D} = 
-  \begin{pmatrix}
         M_D \\
         M_{RS}
        \end{pmatrix} M^{-1}_R
                  \begin{pmatrix}
                   M^T_D & M^T_{RS} 
                  \end{pmatrix} \nonumber \\
    &=& - \begin{pmatrix}
       M_{D} M^{-1}_R M^T_{D}    & M_{D} M^{-1}_R M^T_{RS}  \\
       M_{RS} M^{-1}_R M_{D}          &  M_{RS} M^{-1}_R M^T_{RS}
       \end{pmatrix}
       \label{eq:nuL-nuS}
\end{eqnarray}
Here the heavy right-handed neutrinos are integrated out from the block diagonalized mass formula and can be written simply as,
\begin{eqnarray}
M^{\rm heavy}_{MERS}&\equiv& m_N =
   M_R + \cdots
   \label{eq:nuR}
\end{eqnarray}
The resulting approximate block diagonalized mixing matrix $\mathcal{W}_1$ in terms of masses is given by:
\begin{eqnarray}
\mathcal{W}_1 &=& \begin{pmatrix}
            1- \frac{1}{2} Z Z^\dagger & -\frac{1}{2} Z Y^\dagger & Z  \\
            -\frac{1}{2} Y Z^\dagger & 1- \frac{1}{2} Y Y^\dagger  & Y  \\
            - Z^\dagger   & -Y^\dagger &  1-\frac{1}{2}\left(Z^\dagger Z + Y^\dagger Y\right)
                                 \end{pmatrix}
 \end{eqnarray}
where $Z=M_D M^{-1}_R$ and $Y=M_{RS} M^{-1}_{R}$. Thus, the effective light neutral fermions states $(\nu_L, \nu_S)$ are completely decoupled from the heaviest right-handed neutrinos because of suppressed mixing between light and heavy states.

\noindent \\
{\bf \underline{Second Seesaw Approximation with $\mathcal{W}_2$ for :-}}\\
Upon the application of type-I seesaw procedure first time, we integrated out the right-handed neutrinos as heaviest Majorana particles with mass $M_R$ and got the light active neutrinos $\nu_L$ and sterile neutrinos $\nu_S$ in a block diagonal form as,
\begin{eqnarray}
M^{\rm Eff}_{MERS} &=&-
\begin{pmatrix}
       M_{D} M^{-1}_R M^T_{D}    & M_{D} M^{-1}_R M^T_{RS}  \\
       M_{RS} M^{-1}_R M_{D}          &  M_{RS} M^{-1}_R M^T_{RS}
       \end{pmatrix} = 
         \begin{pmatrix}
              \mathcal{M}^\prime_L& \mathcal{M}^{\prime T}_D  \\
              \mathcal{M}^\prime_{D} & \mathcal{M}^\prime_{R}
             \end{pmatrix} \, , \nonumber \\
 \mbox{where,} && \mathcal{M}^\prime_{L}= -M_{D} M^{-1}_R M^T_{D} \nonumber \\
    &&            \mathcal{M}^\prime_{D}= -M_{RS} M^{-1}_R M^T_{D} \nonumber \\
    && \mathcal{M}^\prime_{R}= -M_{RS} M^{-1}_R M^T_{RS}
       \label{eq:nuL-nuS-b}
\end{eqnarray}
This mass matrix looks exactly like type-I+II seesaw structure and thus the application of type-I+II seesaw approximation is required one more time in eq.(\ref{eq:nuL-nuS}). Applying the second seesaw approximations $| \mathcal{M}^{\prime}_{R} | \gg |\mathcal{M}^{\prime}_{D}| \gg |\mathcal{M}^{\prime}_{L}| $, we get the resulting block diagonalized MERS mass matrix as,
\begin{eqnarray}
&&\mathcal{S}^T \mathcal{M}^{\rm Eff}_{\tiny \rm MERS} \mathcal{S}
     =\begin{pmatrix}
        m_\nu & 0 \\
        0     & m_S
      \end{pmatrix}
 \nonumber \\
    && \mathcal{S}=\begin{pmatrix}
             \sqrt{1-\mathcal{A}\mathcal{A}^{\dagger}} & \mathcal{A}\\
             -\mathcal{A}^{\dagger} & \sqrt{1-\mathcal{A}^{\dagger}\mathcal{A}}
            \end{pmatrix}
        =\begin{pmatrix}
         1- \frac{1}{2} X X^\dagger & X \\
            -X^\dagger & 1- \frac{1}{2} X^\dagger X
         \end{pmatrix}
\end{eqnarray}

\begin{eqnarray}
&&\mathcal{W}^T_2 \mathcal{M}^{\rm BD}_{\tiny \rm MERS} \mathcal{W}_2=\mathcal{M}^{\rm dia}_{\tiny \rm MERS} \nonumber \\
    && \mathcal{M}^{\rm diag}_{\tiny \rm MERS} = \begin{pmatrix}
                                        m_\nu  &  0        & 0\\
                                        0      &  m_S      & 0 \\
                                        0      &  0        & m_N
                                       \end{pmatrix} \nonumber \\
   && \mathcal{W}_2 = 
             \begin{pmatrix}
              \mathcal{S}  & {\bf 0}_{6 \times 3} \\
              {\bf 0}_{3 \times 6} & {\bf 1}_{3 \times 3}
             \end{pmatrix} \nonumber \\
  &&\hspace*{1cm} =\begin{pmatrix}
    \sqrt{1-\mathcal{A}\mathcal{A}^{\dagger}} & \mathcal{A}   &  {\bf 0}_{3 \times 3} \\
-\mathcal{A}^{\dagger} & \sqrt{1-\mathcal{A}^{\dagger}\mathcal{A}} & {\bf 0}_{3 \times 3}\\
  {\bf 0}_{3 \times 3}  & {\bf 0}_{3 \times 3} & {\bf 1}_{3 \times 3} 
            \end{pmatrix}
            =\begin{pmatrix}
    1- \frac{1}{2} X X^\dagger & X  &  {\bf 0}_{3 \times 3} \\
    -X^\dagger & 1- \frac{1}{2} X^\dagger X & {\bf 0}_{3 \times 3}\\
  {\bf 0}_{3 \times 3}  & {\bf 0}_{3 \times 3} & {\bf 1}_{3 \times 3} 
            \end{pmatrix}
\end{eqnarray}

The resulting mass formulas for light active neutrinos, sterile neutrinos and heavy neutrinos, respectively are given as
\begin{eqnarray}
m_{\nu} &=& \mathcal{M}^\prime_{L}- \mathcal{M}^{\prime T}_{D} \mathcal{M}^{\prime^{-1}}_{R} \mathcal{M}^{\prime}_{D}\nonumber \\
&=&-M_{D} M^{-1}_R M^T_{D} - 
     \big(-M_{D} M^{-1}_R M^T_{RS} \big) 
     \cdot \big(-M_{RS} M^{-1}_R M^T_{RS} \big)^{-1} 
     \cdot \big(-M_{RS} M^{-1}_R M^T_{D} \big) \nonumber \\
&=&-M_{D} M^{-1}_R M^T_{D} + M_{D} M^{-1}_R M^T_{D} \nonumber \\
&=& 0        \label{eq:mnu1} \\
m_{S} &=& -M_{RS} M^{-1}_R M^T_{RS}  \\
   \label{eq:mS1}
m_{N} &=& M_R 
   \label{eq:mN1}
\end{eqnarray}
It is now established that type-I seesaw contribution gets completely canceled out at leading order while sterile and heavy right-handed neutrinos have block diagonal matrices. To completely diagonalize these matrices, we need another unitary mixing matrix as
\begin{eqnarray}
 \mathcal{U}_{9 \times 9} = \begin{pmatrix}
   {\bf 1}_{3 \times 3}  &  {\bf 0}_{3 \times 3}  & {\bf 0}_{3 \times 3}  \\
   {\bf 0}_{3 \times 3}  &  {U_S}_{3 \times 3}    & {\bf 0}_{3 \times 3}  \\
   {\bf 0}_{3 \times 3}  &  {\bf 0}_{3 \times 3}  & {U_N}_{3 \times 3}
\end{pmatrix} 
\end{eqnarray}

So we arrive at the complete mixing matrix as,
\begin{eqnarray}
 \mathcal{V}_{9 \times 9} &=& \mathcal{W}_1 \cdot \mathcal{W}_2 \cdot \mathcal{U}_{9 \times 9} 
 \nonumber 
\end{eqnarray}

\subsection{Radiative contribution to light neutrino masses in MERS}
\label{section:Radiative}
The interesting feature of the model is that the heaviest right-handed neutrinos generate light active neutrino masses at one loop level via exchange of SM $W$ and $Z$ gauge bosons as shown in figure~\ref{fig:figure1}. The one loop contribution to light neutrino masses in the limit $\|M_R\| \gg
\|M_D\|,\|M_{RS}\|$ is given by~\cite{Lopez-Pavon:2015cga}
\begin{eqnarray}
m^{1-{\rm loop}}_\nu \equiv \Delta M &\simeq & M_D \frac{\alpha_W}{16\pi
  m_W^2} M_R\left[\frac{m_H^2}{M_R^2-m_H^2{\bf 1}_3}\ln
  \left(\frac{M_R^2}{m_H^2}\right) +   
	\frac{3m_Z^2}{ M_R^2-m_Z^2{\bf 1}_3}\ln\left(\frac{
          M_R^2}{m_Z^2}\right)\right] M^T_D\nonumber\\ 
& \simeq &  
M_D M_R^{-1} x_R\, f(x_R) M_D^{T}
\end{eqnarray}
where the one-loop function $f(x_R)$ is defined as 
\begin{eqnarray}
f(x_R) =
\frac{\alpha_W}{16\pi}\left[\frac{x_H}{x_R-x_H}\ln\left(\frac{x_R}{x_H}\right)
  + \frac{3x_Z}{x_R-x_Z}\ln\left(\frac{x_R}{x_Z}\right) \right] 
\end{eqnarray} 
with   $x_R   \equiv\hat{M}_R^2/m_W^2$,   $x_H\equiv   m_H^2/m_W^2$,
$x_Z\equiv m_Z^2/m_W^2$, $\hat{M}_R$ as diagonal matrix.

This loop contribution can be added in the effective mass matrix 
\begin{eqnarray}
M^{\rm Eff}_{MERS} &=&-
\begin{pmatrix}
       M_{D} M^{-1}_R M^T_{D} -\Delta M   & M_{D} M^{-1}_R M^T_{RS}  \\
       M_{RS} M^{-1}_R M_{D}          &  M_{RS} M^{-1}_R M^T_{RS}
       \end{pmatrix} \nonumber \\
      && = -\begin{pmatrix}
       M_{D} M^{-1}_R \bigg({\bf 1}_{3} - x_R f\left(x_R \right) \bigg) M^T_{D}  & M_{D} M^{-1}_R M^T_{RS}  \\
       M_{RS} M^{-1}_R M_{D}          &  M_{RS} M^{-1}_R M^T_{RS}
       \end{pmatrix}
\end{eqnarray}
The already discussed block diagonalization procedure can be further applied to diagonalize light active neutrino masses. With this radiative contribution of light neutrino masses in $M^{\rm Eff}_{MERS}$, the modified masses and mixings are 

\begin{eqnarray}
&&\mathcal{W}^T_2 \mathcal{M}^{\rm BD}_{\tiny \rm MERS} \mathcal{W}_2=\mathcal{M}^{\rm dia}_{\tiny \rm MERS} \nonumber \\
    && \mathcal{M}^{\rm diag}_{\tiny \rm MERS} = \begin{pmatrix}
                                        m_\nu  &  0        & 0\\
                                        0      &  m_S      & 0 \\
                                        0      &  0        & m_N
                                       \end{pmatrix} \nonumber \\
   && \mathcal{W}_2 = 
             \begin{pmatrix}
              \mathcal{S}  & {\bf 0}_{6 \times 3} \\
              {\bf 0}_{3 \times 6} & {\bf 1}_{3 \times 3}
             \end{pmatrix} \nonumber \\
  &&\hspace*{1cm} =\begin{pmatrix}
    \sqrt{1-\mathcal{A}\mathcal{A}^{\dagger}} & \mathcal{A}   &  {\bf 0}_{3 \times 3} \\
-\mathcal{A}^{\dagger} & \sqrt{1-\mathcal{A}^{\dagger}\mathcal{A}} & {\bf 0}_{3 \times 3}\\
  {\bf 0}_{3 \times 3}  & {\bf 0}_{3 \times 3} & {\bf 1}_{3 \times 3} 
            \end{pmatrix}
            =\begin{pmatrix}
    1- \frac{1}{2} X X^\dagger & X  &  {\bf 0}_{3 \times 3} \\
    -X^\dagger & 1- \frac{1}{2} X^\dagger X & {\bf 0}_{3 \times 3}\\
  {\bf 0}_{3 \times 3}  & {\bf 0}_{3 \times 3} & {\bf 1}_{3 \times 3} 
            \end{pmatrix}
\end{eqnarray}

\noindent So, after the generation of radiative mass for light neutrinos, the resulting mass formulas for light active neutrinos, sterile neutrinos and heavy neutrinos are given by
\begin{eqnarray}
\label{eq:massformula}
m_{\nu} &=& \Delta M =  M_{D} M^{-1}_R x_R f\left(x_R \right) M^T_{D}     \label{eq:mnu2} \\
m_{S} &=& -M_{RS} M^{-1}_R M^T_{RS}  \\
   \label{eq:mS2}
m_{N} &=& M_R 
   \label{eq:mN2}
\end{eqnarray}

\subsection{Complete masses and mixing for MERS}
\begin{eqnarray}
  \label{eq:mformula}
m_{\nu} &\simeq& \left({\bf 1}_3+\zeta\zeta^\dag\right)^{-1/2} \bigg(M_D M_R^{-1} x_R\, f(x_R) M_D^{T} \bigg) \left({\bf
  1}+\zeta^*\zeta^{\sf T}\right)^{-1/2}\; ,\\ 
   \label{eq:m2}
m_{S} &=& M^\prime_{RS} M_R^{-1} M^{\prime T}_{RS} \simeq -M_{RS} M_R^{-1}M^{ T}_{RS}\; \\
m_N &=& M_R 
\end{eqnarray}
\begin{eqnarray}
 \mathcal{V}_{9 \times 9} &=& \mathcal{W}_1 \cdot \mathcal{W}_2 \cdot \mathcal{U}_{9 \times 9} 
 \nonumber \\
\end{eqnarray}

The first block diagonalized mixing matrix is given by
\begin{eqnarray}
&&\mathcal{W}_1 = \begin{pmatrix}
                  1 -\frac{1}{2} Z Z^\dagger   &  -\frac{1}{2} Z Y^\dagger  &  Z \\
                 -\frac{1}{2} Y Z^\dagger  &  1-\frac{1}{2} Y Y^\dagger  & Y  \\
                 -Z^\dagger   & -Y^\dagger  & 1-\frac{1}{2}\left(Z^\dagger Z + Y^\dagger Y \right)
                 \end{pmatrix} \nonumber \\
&&\mathcal{W}_2 = \begin{pmatrix}
                 1 -\frac{1}{2} X X^\dagger & X   & 0 \\
                 -X^\dagger & 1 -\frac{1}{2} X^\dagger X  & 0 \\
                 0    &     0       & 1   
                  \end{pmatrix}
\end{eqnarray}
The unitary matricee which will diagonalize the already block derived mass formulas in order to get the physical masses is given by
\begin{eqnarray}
\mathcal{U} = \begin{pmatrix}
     U_\nu & 0 & 0 \\
     0   & U_S  & 0 \\
     0  &  0 &  U_N
    \end{pmatrix}
\end{eqnarray}
such that 
\begin{eqnarray}
 &&U^\dagger_\nu m_\nu U^*_\nu = \hat{m}_\nu = \mbox{diag}\left(m_{\nu_1}, m_{\nu_2}, m_{\nu_3} \right) \nonumber \\
 &&U^\dagger_S m_S U^*_S = \hat{m}_S = \mbox{diag}\left(m_{s_1}, m_{s_2}, m_{s_3} \right) \nonumber \\
  &&U^\dagger_N m_N U^*_N = \hat{m}_N = \mbox{diag}\left(m_{N_1}, m_{N_2}, m_{N_3} \right)
\end{eqnarray}

The complete $9\times 9$ mixing matrix for the present work is derived as,
\begin{eqnarray}
 \mathcal{V} &=& \mathcal{W}_1 \cdot \mathcal{W}_2 \cdot \mathcal{U} \nonumber \\
    &&\hspace*{-0.5cm}=\begin{pmatrix}
                  1 -\frac{1}{2} Z Z^\dagger   &  -\frac{1}{2} Z Y^\dagger  &  Z \\
                 -\frac{1}{2} Y Z^\dagger  &  1-\frac{1}{2} Y Y^\dagger  & Y  \\
                 -Z^\dagger   & -Y^\dagger  & 1-\frac{1}{2}\left(Z^\dagger Z + Y^\dagger Y \right)
                 \end{pmatrix} \cdot 
                 \begin{pmatrix}
                 1 -\frac{1}{2} X X^\dagger & X   & 0 \\
                 -X^\dagger & 1 -\frac{1}{2} X^\dagger X  & 0 \\
                 0    &     0       & 1   
                  \end{pmatrix} \cdot 
                  \begin{pmatrix}
     U_\nu & 0 & 0 \\
     0   & U_S  & 0 \\
     0  &  0 &  U_N
    \end{pmatrix} \nonumber
\end{eqnarray}

The complete mixing matrix is
\begin{eqnarray}
\label{eq:mixingformula}
\mathcal{V}
&=& \begin{pmatrix}
U_{\nu}\left(1-\frac{1}{2}X X^{\dagger} \right)  & -U_{S}X &  Z U_N\\
   -U_{\nu}X^{\dagger}  & U_S \left(1-\frac{1}{2}X^{\dagger}X \right) & U_{N} Y \\
   \left(Z^\dagger X^\dagger X\right)U_\nu & -U_S Y^{\dagger} & U_N\left(1-\frac{1}{2}Y^{\dagger}Y \right)
\end{pmatrix}
\end{eqnarray}
here $ X = M_{RS} M^{-1}_{R}$, $Y = M_{RS} M^{-1}_{R}$ and $Z=M_D M^{-1}_R$ ,if we consider $ M_D \simeq 10^{-1}MeV - MeV $ ,$ M_{RS} \simeq 1 MeV - 10\,MeV $ and $ M_{R} \simeq 1\,GeV - 200\,GeV $ then we get $X \simeq 0.1 $, $Y\simeq 0.001$, $Z=10^{-5}$ and since $U_{\nu}$,$U_N$and $U_S$ are of the order of $0.9-1.0$ then the matrix elements of $\mathbb{V}$ are comming out to be
\begin{eqnarray}
\begin{pmatrix}
               \mathcal{V}_{\alpha i}^{\nu \nu} & \mathcal{V}_{\alpha j}^{\nu S} & \mathcal{V}_{\alpha k}^{\nu N}\\\mathcal{V}_{\beta i}^{ S \nu} & \mathcal{V}_{\beta j}^{S S} & \mathcal{V}_{\beta k}^{S N}\\\mathcal{V}_{\gamma i}^{N \nu} & \mathcal{V}_{\gamma j}^{N S} & \mathcal{V}_{\gamma k}^{N N}
               \end{pmatrix}  \simeq \begin{pmatrix}
               1       & 0.1     & 10^{-5} \\
               0.1     & 0.91    & 0.001   \\
               0       & 0.001   & 1.0
               \end{pmatrix}
\end{eqnarray} 
that is the sizable contribution to neutrinoless double beta decay only comes from $\mathcal{V}_{\alpha i}^{\nu \nu}$ ,$\mathcal{V}_{\alpha j}^{\nu S}$ and all the other terms are very small
\begin{eqnarray}
\nu_{\alpha} &=&\sum_{i=1,2,3}\mathcal{V}_{\alpha i}^{\nu \nu} \nu_{i} + \mathcal{V}_{\alpha j}^{\nu S} S_j + \mathcal{V}_{\alpha k}^{\nu N} N_k \nonumber \\
\label{eq:ccLagrangian}
\end{eqnarray}

\bibliographystyle{utcaps_mod}
\bibliography{onubb_LR}
\end{document}